\documentclass[aps,floatfix,onecolumn,a4paper,noshowpacs, nofootinbib,superscriptaddress,11pt]{revtex4}

\usepackage{graphicx,float,tikz}
\usepackage[all]{xy}
\usepackage{physics,bm,amsmath,upgreek,bigints}
\usepackage{slashed}

\usepackage{amssymb}
\usepackage{color}
\usepackage{epsfig,bm}		
\usepackage{graphicx,epstopdf}
\usepackage{subfigure,upgreek}
\usepackage{pdfpages}
\usepackage{multirow}
\usepackage{array}
\newcommand{\beq}{\begin{eqnarray}}
\newcommand{\eeq}{\end{eqnarray}}

\newcommand{\clt}{\textcolor{black}}

\usepackage{bookmark,textgreek}
\usepackage{hyperref,color,xcolor}
\hypersetup{hidelinks,colorlinks=true,breaklinks=true,urlcolor= blue}
\hypersetup{%
    colorlinks = true,
    linkcolor  = blue,
    citecolor = cyan,
  }

\usepackage{natbib}
\setcitestyle{square,numbers}
\newcommand{\be}{\begin{equation}}
\newcommand{\ee}{\end{equation}}
\newcommand{\p}{\partial}
\newcommand{\bea}{\begin{eqnarray}}
\newcommand{\eea}{\end{eqnarray}}

\newcommand\orcidroldao{{\href{https://orcid.org/0000-0003-3978-532X}{\orcidicon}}}
\newcommand\orcidpedro{{\href{https://orcid.org/0000-0001-9151-0900}{\orcidicon}}}
\newcommand\orcidbruno{{\href{https://orcid.org/0000-0003-0969-9359}{\orcidicon}}}

\newcommand{\orcidicon}{%
	\begin{tikzpicture}
	\draw[lime, fill=lime] (0,0)
		circle [radius=0.16]
		node[white] {{\fontfamily{qag}\selectfont \tiny ID}};
	\draw[white, fill=white] (-0.0625,0.095)
		circle [radius=0.007];
	\end{tikzpicture}	\hspace{-2mm}
}

\begin{document}

\title{Holographic information measures for spin-$3/2$ $\Delta$ baryons in AdS/QCD}

\author{H. Almeida}
\email{hudson.vinicius@ufabc.edu.br}
\affiliation{Federal University of ABC, Center of Physics, Santo Andr\'e, 09580-210, Brazil}

\author{R. da Rocha\orcidroldao\!\!}
\email{roldao.rocha@ufabc.edu.br}
\affiliation{Federal University of ABC, Center of Mathematics, Santo Andr\'e, 09580-210, Brazil}
\author{P. H. O. Silva\orcidpedro\!\!}
\email{silva.pedro@ufabc.edu.br}
\affiliation{Federal University of ABC, Center of Physics, Santo Andr\'e, 09580-210, Brazil}

\author{B. Toniato\orcidbruno\!\!}
\email{bruno.toniato@ufabc.edu.br}
\affiliation{Federal University of ABC, Center of Physics, Santo Andr\'e, 09580-210, Brazil}

\begin{abstract}
 Spin-$3/2$ $\Delta$ baryon resonances are investigated within AdS/QCD, using Rarita--Schwinger fields. The differential configurational entropy (DCE) and differential configurational complexity (DCC) associated with their bulk energy densities are computed. It yields Regge-like trajectories relating configurational information measures to the radial excitation number and the experimental mass spectrum of the $\Delta$ baryons. We then extrapolate the spectrum of heavier $\Delta$ baryon resonances beyond currently established states in the PDG, also comparing them with states in PDG omitted from the summary table. Our results support a relevant interplay among holographic QCD dynamics, configurational information entropy, and baryon spectroscopy in strongly coupled QCD.
\end{abstract}

\maketitle
\section{Introduction}

The AdS/CFT correspondence~\cite{Maldacena:1997re,Witten:1998qj, Gubser:1998bc,Aharony:1999ti} provides a duality between a conformal field theory in four dimensions and a gravitational theory in a five-dimensional anti-de Sitter (AdS) space. Such correspondence, an example of the so-called holographic principle, has motivated the construction of effective models, one of them known as AdS/QCD, which captures essential features of low-energy QCD through gravity in AdS bulk~\cite{Erlich:2005qh,Karch:2006pv}. AdS/QCD successfully reproduces meson and baryon spectra, hadronic form factors, and other nonperturbative phenomena using semiclassical gravitational duals. AdS/QCD targets the low-energy regime of QCD, where quarks and gluons are confined, so hadronic properties must be described using intrinsically strong-coupling dynamics~\cite{Csaki,DaRold:2005mxj,dePaula:2009za}. The soft-wall AdS/QCD model describes key features of QCD, such as confinement, phenomenological Regge trajectories, and asymptotic freedom~\cite{Gursoy:2007cb,Ballon-Bayona:2017sxa,Ballon-Bayona:2017bwk,FolcoCapossoli:2019imm}. This is achieved by introducing a scalar dilaton field in the AdS bulk, which contributes to both the chiral and gluon condensates, allowing linear confinement and chiral symmetry breaking to manifest~\cite{Gherghetta:2009ac,dePaula:2008fp,Colangelo:2008us}. The soft-wall AdS/QCD model can also be deformed to account for the description of heavy-flavor hadrons~\cite{Rinaldi:2024fgx,MartinContreras:2020cyg,Boschi-Filho:2002xih,Braga:2018hjt,Karapetyan:2021ufz,BallonBayona:2007vp,Diles:2025xot} and hadronic matter immersed in a magnetic field~\cite{Jena:2024cqs,Dudal:2014jfa,Dudal:2018rki}.

The Rarita–Schwinger field, describing spin-$3/2$ particles, plays a central role in supersymmetric theories of gravity and their holographic realizations. Within the AdS/CFT correspondence, it represents the bulk dual of a spin-$3/2$ operator in the boundary theory, typically identified with the supercurrent~\cite{Volovich:1998tj,Koshelev:1998tu,Matlock:1999fy}. Its dynamics plays a central role in clarifying how local supersymmetry in the bulk gravitational description is encoded holographically in the boundary theory. In supergravity, the Rarita--Schwinger field corresponds to the gravitino~\cite{Henneaux:1985kr,Dantas:2015dca,Gomes:2023qkj,Kirsch:2006he,Benakli:2023aes}.  
Later works extended these analyses to fluid/gravity dualities, black hole superpartners, and fermionic wigs that emulate black hole bosonic hair~\cite{Gentile:2012jm,Shaukat:2009hp}. 

Baryons are essential components of QCD. They have been well explored in AdS/QCD~\cite{Hong:2006ta, Fang:2016uer, Pomarol:2008aa}, by extending the well-established meson models to describe baryons as well, including effects from chiral symmetry breaking~\cite{Ahn:2009px}. Further works used AdS/QCD at finite temperature to explore temperature effects on the baryon spectrum~\cite{Guo:2024nrf,daRocha:2025gcz, Gutsche:2019pls, Wang:2015osq}. In the nonperturbative regime of QCD, spin-$3/2$ baryonic resonances, such as the $\Delta$ baryons, provide an important testing ground for models of confinement and chiral symmetry breaking~\cite{Capstick:1986bm}. Embedding the Rarita–Schwinger action in an AdS/QCD framework provides a holographic description of $\Delta$ baryons and other spin-$3/2$ resonances, from which one can determine their mass spectrum, couplings, and form factors~\cite{CoronadoVillalobos:2015mns,Meert:2018qzk}. The $\Delta$ baryon resonances also play a prominent role in studying neutron stars, as Ref.~\cite{Marquez:2022gmu} showed that compact stars, the so-called deltic stars, present a significant amount of $\Delta$ baryons in the nucleonic matter at their center. In addition, Ref.~\cite{Parmar:2025csx} showed that at intermediate baryon densities, the onset of $\Delta$ baryons reduces pressure values and that $\Delta$ baryons yield stable astrophysical neutron-star configurations with maximum masses around twice the solar mass.

Recent studies have used configurational information measures (CIMs) to analyze the  structure of hadronic states~\cite{Gleiser:2011di,Gleiser:2012tu,Bernardini:2016qit}. CIMs quantify the informational content of localized field configurations and correlate them with the stability and dominance of physical states. Combining the Rarita--Schwinger formalism within AdS/QCD to CIMs provides a novel perspective on the holographic investigation of spin-$3/2$ $\Delta$ baryons and the mass spectrum of heavier spin-$3/2$ $\Delta$ baryon resonances. One of the CIMs is the differential configurational entropy (DCE), which provides an information-theoretic measure in QCD~\cite{Gleiser:2018jpd} and has proven valuable for investigating hadronic properties within holographic QCD models~\cite{Bernardini:2018uuy,Karapetyan:2018oye,Colangelo:2018mrt,Braga:2023qee,MartinContreras:2022lxl,Braga:2020opg,MartinContreras:2023eft,MartinContreras:2023oqs}. Refs.~\cite{Dudal:2018ztm,daRocha:2021xwq} also addressed phenomenological features of information theory in AdS/QCD, using holographic entanglement entropy. 

Using the Rarita--Schwinger formalism for spin-$3/2$ fields, we construct the holographic duals of $\Delta$ baryons and analyze their spectral profiles. The information content of these modes is quantified through CIMs derived from their probability distributions, providing insight into the structure of the corresponding boundary operators. The DCE and the differential configurational
complexity (DCC) can capture essential features of the $\Delta$ baryon mass spectrum, and offer a robust tool for characterizing heavier $\Delta$ baryon resonances in strongly coupled gauge theories. This work establishes a direct link between Rarita--Schwinger dynamics in AdS and the information-theoretic structure of spin-$3/2$ baryons in the boundary theory.
 
The main goal of this work is to analyze the information-theoretic properties of spin-$3/2$ $\Delta$ baryons coming from a holographic model for the Rarita–Schwinger field. By computing the DCE and the DCC underlying the $\Delta$ baryons, we obtain the mass spectrum of heavier $\Delta$ baryon resonances based on the experimental mass spectrum in PDG~\cite{PDG2024}, establishing then a deeper connection between information theory and phenomenological aspects of holographic QCD. This paper is organized as follows: Section~\ref{section:themodel} introduces the AdS/QCD hard-wall framework for spin-$3/2$ $\Delta$ baryons, describing their holographic realization in terms of five-dimensional Rarita–Schwinger fields and the implementation of chiral symmetry breaking. The resulting equations of motion and boundary conditions determine the $\Delta$ baryon mass spectrum, which is compared with experimental data. Sections~\ref{section:dce} and~\ref{section:dcc}, respectively, show the computation of the DCE and DCC from the holographic energy density of the Rarita–Schwinger fields. Regge-like trajectories are constructed and used to extrapolate the masses of heavier $\Delta$ baryon resonances. Section~\ref{section:conclusion} summarizes the main results and discusses their implications for holographic baryon spectroscopy.

\section{Spin-$3/2$ $\Delta$ baryon resonances in AdS/QCD}\label{section:themodel}

The family of $\Delta$ baryons has isospin values of $I_3=3/2, \,1/2$ and spin $J = 3/2$. $\Delta$ baryons quickly decay via the strong interaction to a nucleon ($N$) and a pion ($\pi$) \cite{PDG2024}. Due to their strong coupling to the $\pi N$ channel, $\Delta$ baryon resonances are essential in modeling the hadronic phase following quark-gluon plasma (QGP) hadronization \cite{Rougemont:2015ona}. After the formation of the QGP phase in relativistic heavy-ion collisions, it cools down and hadronizes into a strongly interacting hadronic phase, in which $\Delta$ baryon resonances are abundantly produced. They dominate $\pi N$ dynamics and significantly shape the final pion and nucleon spectra. The $\Delta/N$ ratio serves as a sensitive probe of the chemical freeze-out temperature and baryon chemical potential in statistical hadronization models~\cite{Andronic2018}, while the reaction $\pi N \leftrightarrow \Delta$ maintains chemical equilibrium until kinetic freeze-out.

To obtain a holographic description of the chiral dynamics of hadrons via AdS/QCD, one must define the bulk content. Despite the many operators that can be defined in QCD, we are only interested in those related to the chiral dynamics. The left- and right-handed currents corresponding to the $SU(2)_L\times SU(2)_R$ chiral flavor symmetry will be associated with the five-dimensional gauge fields $L_M$ and $R_M$, respectively. One also introduces a bulk scalar field $X$, associated with the quark bilinear operator $\bar{q}_Lq_R$, which acts as an order parameter for chiral symmetry breaking. The minimal bulk action that describes the chiral dynamics is given as~\cite{Erlich:2005qh} 
\be
S=\int d^5x\,\, \sqrt{G}\, \Tr\Big[ |DX|^2 + M_5 X^2 - \frac{1}{4g_5}(F_L^2 + F_R^2)\Big]\,,
\ee
where $G$ is the absolute value of the determinant of the AdS$_5$ Poincaré metric 
 \be
ds^2 = \frac{1}{z^2}
\left( \eta_{\mu\nu} dx^\mu dx^\nu + dz^2 \right),
\qquad \epsilon < z < z_{\textsc{m}}\,,
\ee
where $\eta_{\mu\nu} = \mathrm{diag}(-,+,+,+)$. The cut-off at $z\rightarrow0$ is responsible for regulating the bulk action at the UV boundary, while the IR cut-off at $z=z_{\textsc{m}}$ corresponds to the confinement scale of QCD, $\Lambda_{\rm QCD}$.

Solving the classical equation of motion for the bulk scalar field $X(x,z)=X_0(z)\mathbf{1}$, one gets
\be X_0(z)=\frac{1}{2}m_qz+\frac{1}{2}\sigma z^3\,.\ee

The AdS/CFT dictionary states that the coefficient of the non-normalizable modes, $m_q$, corresponds to the source for the boundary operator. So one identifies $m_q$ as the quark mass, which explicitly breaks the chiral symmetry. The coefficient of the normalizable term $\sigma$ corresponds to the vacuum expectation value of the boundary operator, thus being identified as the quark condensate $\sigma=\langle \bar{q}_Lq_R\rangle$, which spontaneously breaks chiral symmetry. Following Ref.~\cite{Erlich:2005qh}, one fixes $z_{\textsc{m}}^{-1}=346$ MeV, $m_q=2.3$ MeV, and $\sigma=(308\,\text{MeV})^3$, which fit well the light-flavor meson sector.

To obtain a holographic description of spin-$3/2$ baryons in the boundary theory, one has to construct an action for the five-dimensional Rarita--Schwinger fields $\Psi_M$ in the bulk as~\cite{Ahn:2009px}
\be 
S = \int d^4x\int^{\xi z_{\textsc{m}}}_\epsilon dz\, \sqrt{G}\, \bar\Psi_A \left( i \Gamma^{ABC} D_B - m_1 G^{AC} - m_2 \Gamma^{AB} G_B{}^{C} \right) \Psi_C \label{RS5D}\,,
\ee
where $\Psi_A=e^M_A\Psi_M$, with the five-dimensional  vielbein $e^M_A$ satisfying $G_{MN}=e^A_Me^B_N\eta_{AB}$, and $\Gamma^A=(\gamma^\mu,-i\gamma^5)$ are the five-dimensional  gamma matrices, with $\gamma$'s in the chiral representation. We also used the notations $\Gamma^{ABC}=\frac{1}{2}(\Gamma^{B}\Gamma^{C}\Gamma^{A}-\Gamma^{A}\Gamma^{C}\Gamma^{B})$ and $\Gamma^{AB}=\frac{1}{2}[\Gamma^{A},\Gamma^{B}]$. 

One may notice that, unlike in the bosonic sector, in the action~(\ref{RS5D}), a slightly different IR-cutoff is used at the bulk coordinate. This description of baryons can be thought of as being a natural extension of the mesonic model. However, to have good agreement with experimental data, the IR-cutoff needs to differ from the one used in the meson sector. We could simply use a different $z_{\textsc{m}}$, but one would encounter some difficulties in studying the meson-baryon coupling \cite{Wang:2015osq}. A simple solution is to adopt a new parameter $\xi$ and the same $z_{\textsc{m}}$ value as the mesonic sector, indicating that baryons have a confinement scale different from mesons.

The covariant derivative is defined as
\be D_M=\p_M-\frac{i}{4}\omega_M^{AB}\Sigma_{AB}-L_M^a(x,z/\xi)t^a\,,\ee
where $\omega_M^{AB}$ is the spin connection and $\Sigma_{AB}=-i\Gamma_{AB}$. Varying the five-dimensional  Rarita--Schwinger action with respect to $\Psi_A$, one gets the equation of motion~\cite{Ahn:2009px}
\be i\Gamma^{ABC}D_B\Psi_C-m_1\Psi^{A}-m_2\Gamma^{AB}\Psi_B\,, \ee
which can be rewritten as
\be i\Gamma^{B}(D_B\Psi_A-D_A\Psi_B)-m_-\Psi_A+\frac{1}{3}m_+\Gamma_{A}\Gamma^{B}\Psi_B=0\,,\ee
where $m_\pm=m_1\pm m_2$. The parameter $m_-$ directly influences the radial profile of the wavefunction and, consequently, observables like form factors and the energy density distribution along the extra dimension $z$.
Being a reducible vector spinor, the Rarita--Schwinger fields contain not only spin-$3/2$ components
but also spin-$1/2$ components as well. The spurious degrees can be eliminated by the Lorentz covariant constraint $ e^M_A\Gamma^A\Psi_M=0,$ 
which, combined with the equations of motion, gives the condition $\p_M\Psi^M=0$.

The five-dimensional Rarita-Schwinger fields have one extra spin-$1/2$ component compared to the four-dimensional ones. This extra spin-$1/2$ can be projected out by choosing $\Psi_z=0$, since there is no extra spinor in the boundary that it can be mapped into \cite{Ahn:2009px}. With these constraints, the equation of motion reduces to
\be(iz\Gamma^A\p_A-2i\Gamma^5-m_-)\Psi_\mu=0\,.\label{eom2}\ee

Baryons in ADS/QCD are described by a pair of Rarita--Schwinger fields $\Psi^M_1$ and $\Psi^M_2$ in the bulk, corresponding to the four-dimensional  baryon operators $\mathcal{O}_L$ and $\mathcal{O}_R$, respectively. Both fields obey the Rarita--Schwinger equation, albeit with $1\leftrightarrow2$ and $m_-\leftrightarrow-m_-$ to be consistent with the chirality of the boundary operators.

Decomposing the bulk fields chiral components $\Psi^A_{iL,R}$, using the four-dimensional Fourier decomposition \cite{Ahn:2009px}, leads us to
\be \Psi^A_{iL,R}(x,z)=\frac1{2\pi}\int d^4x F_{iL,R}(k,z)\psi_{L,R}^A(k)e^{-ik\cdot x},\quad\quad i=1,2\,, \label{fourier2}\ee
where $\psi^A_{L,R}=(1\pm\gamma^5)\psi^A$ and the four-dimensional spinor $\psi^A$ is defined in a way to satisfy the usual Dirac equation
\be \label{sat1}(\slashed{k}-M_\Delta)\psi^A=0\,,\ee
for 
$\slashed{k}=\gamma^\mu k_\mu$, which implies $\slashed{k}\psi^A_{L,R}=M_\Delta\psi^A_{R,L}$. 

Then, substituting these results into Eq.~(\ref{eom2}), one gets the set of eigenvalue systems
\bea
\Big(\p_z-\frac{1}{z}(2+m_-)\Big)F_{iL}&=&-M_\Delta F_{iR}\,,\\
\Big(\p_z-\frac{1}{z}(2-m_-)\Big)F_{iR}&=&M_\Delta F_{iL}\,.
\eea

The normalizable solutions for the nonzero modes are given by
\be F_{1L,R}\sim z^{5/2}J_{m_-\mp1/2}(M_\Delta z)\,.\ee
 
The AdS/CFT prescription gives the relation $|m_-|=\Delta_{3/2}-2$, where $\Delta_{3/2}=9/2$ is the scaling dimension of the composition operator, so we get $|m_-|=5/2$. In order to get only the left-handed component of the Rarita-Schwinger field on the boundary, one should impose the boundary condition $\Psi^M_{1R}(x,\xi z_{\textsc{m}})=0$. This condition eliminates the right-handed component on the boundary and gives the free spectrum of spin-$3/2$  baryons, determined by the zeros of the Bessel function $J_{m_-+1/2}$. Similarly, to get only the right-handed component on the boundary, one should impose the boundary condition $\Psi^M_{2L}(x,\xi z_{\textsc{m}})=0$. However, for $\Psi^M_2$, the mass term flips the sign, so we end up with the same mass spectrum.

The chiral symmetry breaking is taken into account by introducing a coupling between the bulk scalar field and the Rarita--Schwinger fields. The leading coupling is given by a Yukawa coupling term
\be \mathcal{L}=-g_{3/2}\bar{\Psi}_{1M}X_0(z/\xi)^3\Psi_2^M+ h.c. \,, \ee
The equations of motion~(\ref{eom2}) then become
\bea &&(iz\Gamma^A\p_A-2i\Gamma^5-m_-)\Psi_{1\mu}-g_{3/2}X_0^3\Psi_{2\mu}=0\,, \nonumber\\
&&(iz\Gamma^A\p_A-2i\Gamma^5+m_-)\Psi_{2\mu}-g_{3/2}X_0^3\Psi_{1\mu}=0\,, \eea
and decomposing the chiral basis, using the Fourier transform~(\ref{fourier2}), yields the coupled system
\bea
\begin{pmatrix}
    \p_z-\frac{\Delta_+}{z} & -\frac{g_{3/2}X_0^3}{z} \\
    -\frac{g_{3/2}X_0^3}{z} & \p_z-\frac{\Delta_-}{z}
\end{pmatrix}
\begin{pmatrix}
    F_{1L}\\
    F_{2L}
\end{pmatrix}
&=&-M_\Delta
\begin{pmatrix}
    F_{1R}\\
    F_{2R}
\end{pmatrix}\,,\nonumber\\
\begin{pmatrix}
    \p_z-\frac{\Delta_-}{z} & \frac{g_{3/2}X_0^3}{z} \\
    \frac{g_{3/2}X_0^3}{z} & \p_z-\frac{\Delta_+}{z}
\end{pmatrix}
\begin{pmatrix}
    F_{1R}\\
    F_{2R}
\end{pmatrix}
&=&M_\Delta
\begin{pmatrix}
    F_{1L}\\
    F_{2L}
\end{pmatrix}\,,\label{baryonspectrum}
\eea
where $\Delta_{\pm}=2\pm M_\Delta$ and the boundary conditions are
\be\Psi^A_{1R}(x,\epsilon)=\Psi^A_{1R}(x,\xi z_{\textsc{m}})=0=\Psi^A_{2L}(x,\epsilon)=\Psi^A_{2L}(x,\xi z_{\textsc{m}})\,.\ee

\begin{figure}[H]
	\centering
	\includegraphics[width=10cm]{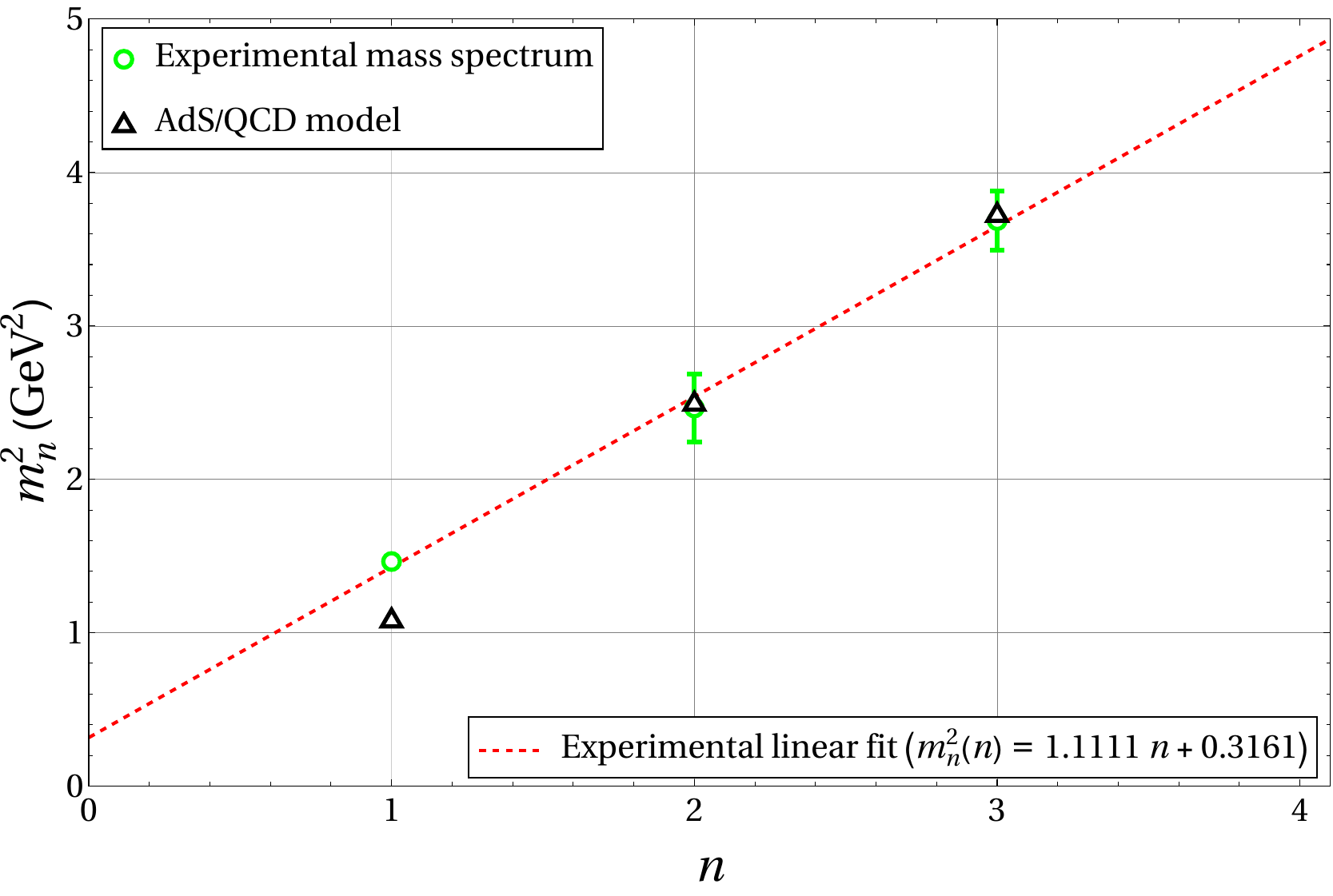}
	\caption{Mass spectra for the $\Delta$ baryon resonances obtained from the experimental
values in PDG \cite{PDG2024} and the AdS/QCD model, for the resonances $\Delta(1232)$, $\Delta(1600)$, and $\Delta(1920)$.}
	\label{fig:dce_n_32}
\end{figure}

Using $\xi=1.5$ and $g_{3/2}=375$, the mass spectrum of $\Delta$ baryon resonances $\Delta(1232)$, $\Delta(1600)$, and $\Delta(1920)$, respectively corresponding to the radial quantum number $n=1,2,3$, is shown in Table~\ref{table:mass}. \textcolor{black}{The parameters $\xi=1.5$ and $g_{3/2}=375$ were chosen phenomenologically to achieve a good agreement between the AdS/QCD predictions and the experimental masses of the lowest-lying $\Delta$ baryon resonances. In the hard-wall AdS/QCD model, the IR cutoff $z_{\textsc{m}}$ is fixed by the meson sector. However, baryons generally require an effective confinement scale that differs slightly from mesons. The parameter $\xi$ rescales the IR cutoff for baryons, $\xi z_{\textsc{m}}$, effectively accounting for their larger spatial extension. The choice $\xi=1.5$ allows the baryon wavefunctions to satisfy the boundary conditions while reproducing the approximate spacing of the radial excitations observed in experiments, consistent with previous studies \cite{Ahn:2009px, Wang:2015osq}. 
On the other hand, the Yukawa coupling $g_{3/2}$ controls the strength of the interaction between the bulk scalar field $X_0(z)$ and the Rarita--Schwinger fields, which encodes the effects of chiral symmetry breaking. Its value directly influences the mass splitting and the radial profile of the baryon modes. By setting $g_{3/2}=375$, the model reproduces the mass of the $\Delta(1232)$ resonance and yields reasonable predictions for higher radial excitations, such as $\Delta(1600)$ and $\Delta(1920)$. Smaller values of $g_{3/2}$ would underestimate the mass gaps, while larger values would overestimate them. Therefore, the chosen values of $\xi$ and $g_{3/2}$ provide a physically motivated and practical calibration of the model parameters to the experimental baryon spectrum.} 
\begin{table}[H]
\begin{center}
\begin{tabular}{||c|c|c|c||}
\hline\hline
        $\quad n \quad$ &\; State \;&\; $M_\text{exp}$ (MeV) \;&\; $M_\text{AdS/QCD}$ (MeV)\; \\\hline\hline \hline
        1 &\; $\Delta(1232)$ \;& $1210\pm10$ & $1038$\\ \hline
        2 & $\Delta(1600)$ & $1570\pm70$ & $1578$ \\ \hline
        3 & $\Delta(1920)$ & $1920\pm50$ & $1929$ \\ \hline\hline 
\end{tabular}
\caption{Mass spectrum of the $\Delta$ baryon family. Column three  depicts the experimental masses from PDG \cite{PDG2024} and column four illustrates the mass spectrum~(\ref{baryonspectrum}) obtained from AdS/QCD.} 
\label{table:mass}
\end{center}
\end{table}
The linear Regge trajectory that interpolates the experimental values of the squared mass as a function of the radial quantum number $n$ can be written as
\begin{flalign}
    m_n^2(n) =  1.1111 n + 0.3161\,.
    \label{schrodingerlike}
\end{flalign}
and it is plotted in Fig.~\ref{fig:dce_n_32}, together with the mass spectrum~(\ref{baryonspectrum}) obtained from AdS/QCD. 

\section{DCE of Spin-$3/2$ $\Delta$ baryon resonances in AdS/QCD}\label{section:dce}
{\color{black}
The DCE quantifies the information content of a physical system and can be introduced if one considers the
correlations of fluctuations of a localized, square-integrable scalar field that characterizes the system, such as the energy density $\rho(x^A)$, where $x^A=(x^\mu,z)$ denote the AdS bulk coordinates, with $A=0,1,2,3,5$ and $\mu=0,1,2,3$. Small perturbations may induce fluctuations in the energy density.
The DCE can be derived from the energy density of the baryonic families, given by the temporal component of the energy--momentum tensor,
\begin{flalign}
    \tau_{00}(x^A)= \frac{2}{\sqrt{g}}\left[ \frac{\partial(\sqrt{g}\mathcal{L})}{\partial g^{00}} - \frac{\partial}{\partial x^\sigma}\left(\frac{\partial(\sqrt{g}\mathcal{L})}{\partial\left(\frac{\partial g^{00}}{\partial x^\sigma}\right)}\right)
    \right]\,,
\end{flalign}
where $\mathcal{L}$ denotes the Lagrangian density defined by the integrand of the action in Eq.~(\ref{RS5D}) that describes the baryons.
However, the baryonic configuration profile must be analyzed in momentum space, which requires performing the Fourier transform
\begin{flalign}
    \langle \tau_{00}(\clt{k_A})\rangle = \frac{1}{(2\pi)^{5}} 
    \int_{\mathrm{AdS}}
    \,\langle \tau_{00}\rangle(x^\mu,z)e^{-ik_A x^A}\,{\sqrt{g}}\,{\rm d}^4x\,dz\,,
    \label{fourier}
\end{flalign}
\clt{where $k_A=(k_\mu,k_z)$ is the AdS bulk wave momentum associated with the coordinates $x^A$. It is important to stress that the four-dimensional spacetime wave momentum can be written, in natural units, as $k_\mu=(\omega,k_i)$, for $i=1,2,3$, where $k_i$ represent the spatial components of the wave momentum and $\omega$ denotes the wave frequency.}
The modal fraction measures the relative contribution of each mode $k_A$ to the full system, and it is defined as \cite{Gleiser:2018jpd}
\begin{flalign}
   \langle \boldsymbol{\tau}_{00}(k_A)\rangle = \frac{\abs{\langle\tau_{00}(k_A)\rangle}^2}{\displaystyle\int_{\mathrm{AdS}}\,\abs{\langle\tau_{00}({k_A})\rangle}^2 \,\sqrt{g}\,d{k_A}}\,,
    \label{dce_modal}
\end{flalign}

\clt{The modal fraction quantifies the way in which wave modes contribute to the power spectral density associated with the energy density. Since the total energy is finite, the power spectral density associated with a mode of momentum $k_A$, contained in the spectral Lebesgue measure $dk_A$, reads 
$P(k_A\,|\,dk_A)\sim |\langle \tau_{00}\rangle(\clt{k_A})|^2\,dk_A$ \cite{Gleiser:2018jpd}, representing  the spectral energy distribution within  $dk_A$. It is worth noting that the power spectral density is proportional to the Fourier transform of the two-point correlator
\begin{eqnarray}
    G(x^A) = \int_{\rm AdS} \langle \tau_{00}(x^A)\rangle\langle\tau_{00}(\tilde{x}^A+x^A)\rangle \sqrt{g}\,d\tilde{x}^A\,,
\end{eqnarray}
which characterizes the DCE as Shannon information entropy of correlations. It also measures the manner in which the energy density fluctuates.}

The DCE, which quantifies the amount of information stored in the $\Delta$ baryons, is defined by
\begin{flalign}
    \mathrm{DCE} = -\int_{\mathrm{AdS}} \check{\boldsymbol{\tau}}_{00} ({k_A}) \log\check{\boldsymbol{\tau}}_{00}({k_A})\,\sqrt{g}\,dk_A\,, 
    \label{dce}
\end{flalign}
where $\langle\check{\boldsymbol{\tau}}_{00}({k_A})\rangle=\langle\boldsymbol{\tau}_{00}({k_A})\rangle/\langle\boldsymbol{\tau}_{00}^\textsc{max}({k_A})\rangle$, and $\langle\boldsymbol{\tau}_{00}^\textsc{max}({k_A})\rangle$ denotes the maximum value of $\langle\boldsymbol{\tau}_{00}({k_A})\rangle$ over the integration domain.
The relevant integrations can be carried out along the AdS bulk away from the boundary \cite{dePaula:2009za}.
For the baryons considered here, the energy density at finite temperature is given by \cite{Ferreira:2020iry}.

The index $M=5$, associated with $x^5\equiv z$, is fixed in Eq.~(\ref{fourier}), whereas $N=5$ will be regarded in Eqs.~(\ref{dce_modal}) and~(\ref{dce}) to estimate the DCE, since a Kaluza--Klein splitting has been taken into account already. In fact, the AdS boundary has codimension one with respect to the AdS bulk, and the $z$ coordinate is the energy scale of QCD (see, e.g.,  Refs.~\cite{Aharony:1999ti,Pirner:2009gr}). Therefore, Eqs.~(\ref{fourier}) to~(\ref{dce}) can be integrated out of the AdS boundary, across the AdS bulk. 

For the Rarita--Schwinger field in AdS$_5$, described by the action~(\ref{RS5D}), with vielbein $e^a{}_M = z^{-1}\delta^a{}_M$ and $\Gamma^M = z\Gamma^a \delta_a{}^M$, 
the stress-energy tensor reads 
\begin{equation}
\langle\tau_{MN}\rangle=\frac{2}{\sqrt G}\frac{\delta S}{\delta G^{MN}}\,,
\end{equation}
and the energy density simplifies to
\begin{align}
\langle \tau_{00}\rangle &=
2 i z\,\partial_0\!\left(
\bar\Psi_A\,\Gamma^{[A0B]}\Psi_B
+ \bar\Psi_0\,\Gamma^{B}\Psi_B
\right)- \frac{i}{z}\,
\bar\Psi_A\,\Gamma^{[A0B]}\Gamma_{0z}\Psi_B
- \frac{i}{2z}\,
\bar\Psi_0\,\Gamma^{B}\Gamma_{0z}\Psi_B \,.
\end{align}

The four-dimensional Rarita--Schwinger field $u_A(k)$ satisfies
Eq.~(\ref{sat1}).
Normalization yields 
\begin{equation}
\bar u_\mu^{(r)}(k) u_\nu^{(s)}(k)
= - (\slashed{k}+M_\Delta)
\left[
g_{\mu\nu}
- \frac{1}{3}\gamma_\mu\gamma_\nu
- \frac{1}{3M_\Delta}(\gamma_\mu k_\nu - \gamma_\nu k_\mu)
- \frac{2}{3M^2_\Delta}k_\mu k_\nu
\right]\delta^{rs}\,,
\end{equation}
where 
a boost in direction of $\vec{k}$ gives
\begin{equation}
u_\sigma(k)=
\sqrt{\frac{E+M_\Delta}{2M_\Delta}}
\begin{pmatrix}
\chi_\sigma \\
\frac{\vec{\sigma}\cdot\vec{k}}{E+M}\chi_\sigma
\end{pmatrix},
\qquad
E=\sqrt{k^2+M_\Delta^2}\,.
\label{posD}
\end{equation} 

Eq.~(\ref{posD}) shows the positive energy Dirac spinors, whereas the spin-$1$ vectors are given by 
\begin{subequations}
\begin{eqnarray}
\epsilon_\mu(+1) &= \displaystyle-\frac{1}{\sqrt{2}}(0,1,i,0)\,,\\
\epsilon_\mu(0)  &= (0,0,0,1)\,,\\
\epsilon_\mu(-1) &= \displaystyle\frac{1}{\sqrt{2}}(0,1,-i,0)\,.
\end{eqnarray}
\end{subequations}

Therefore, the spin-$3/2$ rest-frame states read
\begin{subequations}
\begin{align}
u_\mu(k,+3/2) &= \epsilon_\mu(+1) u_{+1/2}\,,\\
u_\mu(k,+1/2) &= \frac{2}{3}\epsilon_\mu(0)u_{+1/2}
               + \frac{1}{3}\epsilon_\mu(+1)u_{-1/2}\,,\\
u_\mu(k,-1/2) &= \frac{1}{3}\epsilon_\mu(-1)u_{+1/2}
               + \frac{2}{3}\epsilon_\mu(0)u_{-1/2}\,,\\
u_\mu(k,-3/2) &= \epsilon_\mu(-1)u_{-1/2}\,.
\end{align}
\end{subequations}

Polarization vectors transform with the same Lorentz boost.  
The boosted spin-$3/2$ field reads
\begin{equation}
u_\mu(k,s)
=
\sum_{\lambda,\sigma}
\langle 1\,\lambda;\,{\tfrac12}\,\sigma\mid {\tfrac32}\,s\rangle
\,\epsilon_\mu(\lambda)(k) u_\sigma(k)\,.
\end{equation}

Therefore, the Fourier-transformed energy density reads 
\begin{eqnarray}
\langle k_2,s'| \tau_{00}(k,z)|k_1,s\rangle
=
(2\pi)^4\delta^{(4)}(k-(k_1-k_2))
\,f_n^2(z)\,\bar u_A^{(s')}(k_2) M^{AB} u_B^{(s)}(k_1)\,,
\end{eqnarray}
where
\begin{equation}
M^{AB}
=
2z\,\Gamma^{[A0B]}
+ 2z\,\delta_{A0}\Gamma^B
- \frac{i}{z}\Gamma^{[A0B]}\Gamma_{0z}
- \frac{i}{2z}\delta_{A0}\Gamma^{B}\Gamma_{0z}\,.
\end{equation}

Thus, denoting $\langle \tau_{00}\rangle \equiv \langle k_2,s'|\widetilde \tau_{00}(k,z)|k_1,s\rangle$ yields the energy density necessary to compute the DCE as
\begin{eqnarray}
\!\!\!\!\!\!\!\!\!\!\!\!\langle \tau_{00}(k,z)\rangle &\!=\!&
(2\pi)^4\delta^{(4)}(k-(k_2-k_1))
\bar u_A^{(s')}(k_2)
\nonumber\\
&&\times
\Big[
2 z^6 J_\nu^2(kz)\big(\Gamma^{[A0B]}\!+\!\delta^{A}_{0}\Gamma^B\big)
\!-\! i z^4 J_\nu^2(kz)\Gamma^{[A0B]}\Gamma_{0z}
\!-\! \frac{i}{2}z^4 J_\nu^2(kz)\delta^{A}_{0}\Gamma^B\Gamma_{0z}
\Big]
u_B^{(s)}(k_1).\label{t000}
\end{eqnarray}

Eq.~(\ref{t000}) encodes the radial distribution of the energy density carried by spin-$3/2$ baryons in the AdS background and shows how the Rarita--Schwinger field encodes both the vector and spinor components of the $\Delta$ baryon resonances. 

The holographic energy density~(\ref{t000}) provides the starting point for
computing the DCE of spin-$3/2$ $\Delta$ baryon resonances. The dependence on $z$
encodes the shape and nontrivial structure of Rarita--Schwinger modes in AdS, allowing a
quantitative analysis of the spatial information content carried by $\Delta$
baryons. 
To explicitly compute the DCE, numerical integration can be carried out. First, Eq.~(\ref{fourier}) can be rewritten, after performing a change of variables from $z\in[0,+\infty)$ to ${\scalebox{.93}{$\mathfrak{z}$}}\in[0,1]$, as an integral over $[0,1]$, rather than over the entire real line:
\begin{eqnarray}\label{fou1}
    \langle \tau_{00}(k)\rangle &=& \frac{1}{\sqrt{2\pi}} \int_{0}^{+\infty}\langle\tau_{00}(z)\rangle e^{-ikz}dz\nonumber\\
    &=& \frac{1}{\sqrt{2\pi}}\int _{0}^{1}\left\langle\tau_{00} \left({\frac{{\scalebox{.93}{$\mathfrak{z}$}}}{1-{\scalebox{.93}{$\mathfrak{z}$}}}}\right)\right\rangle
    \exp\left[-ik\left({\frac{{\scalebox{.93}{$\mathfrak{z}$}}}{1-{\scalebox{.93}{$\mathfrak{z}$}}}}\right)
   \right]\frac{1}{\left({\scalebox{.93}{$\mathfrak{z}$}-1}\right)^{2}}\,d{\scalebox{.93}{$\mathfrak{z}$}}\,.
\end{eqnarray}

It should also be emphasized that the fluctuation wave functions are introduced mode by mode by solving the corresponding fluctuation equations, with all other fields consistently set to zero. In particular, only the radial wave functions associated with the fluctuations are retained, while the plane-wave factors are omitted. Moreover, when substituting the plane-wave solutions, the gauge fields are required to be real.

In a similar manner, the integration in Eq.~(\ref{dce}) reads
\begin{flalign}
    \text{DCE} = -\int_0^1 \left\langle\check{\boldsymbol{\tau}}_{00}\left({\frac{{\scalebox{.93}{$\mathfrak{z}$}}}{1-{\scalebox{.93}{$\mathfrak{z}$}}}}\right)\right\rangle
    \log\left\langle\check{\boldsymbol{\tau}}_{00}\left({\frac{{\scalebox{.93}{$\mathfrak{z}$}}}{1-{\scalebox{.93}{$\mathfrak{z}$}}}}\right)\right\rangle
    \frac{1}{\left({\scalebox{.93}{$\mathfrak{z}$}-1}\right)^{2}}\,d{\scalebox{.93}{$\mathfrak{z}$}}\,,
    \label{dce1}
\end{flalign}

The integration over $z\in[0,+\infty)$ is effectively truncated by the presence of the hard-wall cutoff $z_{\textsc{m}}$. Consequently, for numerical purposes, the upper integration limit in Eqs.~(\ref{fourier})--(\ref{dce}) can be replaced by the cutoff scale $z_{\textsc{m}}$. Employing the Newton--Cotes quadrature method, the DCE can be evaluated numerically, with the composite iterated trapezoidal rule. For $32768\;(=2^{15})$ grid points, the numerical error is lower than $10^{-7}$, with each DCE value in Eq.~(\ref{dce1}) requiring between 3.2 and 4.0 minutes of computation on an 8-core 4.8~GHz i9 processor running OsX Tahoe. When the Newton--Cotes quadrature is  implemented with Simpson’s rule,  convergent results have numerical errors below $10^{-8}$. For Boole’s rule, numerical errors are within $10^{-9}$, supporting the robustness of the numerical protocol. Since the DCE in Eq.~(\ref{dce1}) is evaluated for the three $\Delta$ baryon resonances, the numerical analysis, although repetitive, remains straightforwardly manageable.

Using the energy density~(\ref{t000}), the DCE values calculated for the $\Delta$ baryon resonances are presented in Table \ref{table:dce}.
\begin{table}[H]
\begin{center}
\begin{tabular}{||c|c|c||}
\hline\hline
        $\quad n \quad$ &\; State \;&\; DCE (nat) \;\\\hline\hline \hline
        1 &\; $\Delta(1232)$ \;& $39.720$ \\ \hline
        2 & $\Delta(1600)$ & $47.511$ \\ \hline
        3 & $\Delta(1920)$ & $58.552$   \\ \hline\hline 
\end{tabular}
\caption{DCE of the radial resonances of the $\Delta$ baryon family.} 
\label{table:dce}
\end{center}
\end{table}

The first type of Regge-like trajectory is obtained by interpolating the calculated DCE values and expressing them as a function of the radial quantum number $n$.
According to the interpolation theorem, the DCE for the three known resonances can be fitted by a second-order polynomial given by
\begin{flalign}
    \text{DCE}_\Delta (n) = 1.625 n^2 + 2.916 n + 35.179\,,
    \label{dcen}
\end{flalign}
with a root-mean-square deviation (RMSD) within $10^{-14}$ \%.
From this expression, it is possible to compute the DCE for higher excited states, i.e., for larger $n$, as presented in Table \ref{table:extdce}. Fig. \ref{fig:dcen} shows this interpolation.

\begin{table}[H]
\begin{center}
\begin{tabular}{||c|c|c||}
\hline\hline
        $\quad n \quad$ &\; State \;&\; DCE (nat) \;\\\hline\hline \hline
        1 &\; $\Delta(1232)$ \;& $39.720$ \\ \hline
        2 & $\Delta(1600)$ & $47.511$ \\ \hline
        3 & $\Delta(1920)$ & $58.552$   \\ \hline
        4 & $\Delta_4^\star$ & $72.843$   \\ \hline
        5 & $\Delta_5^\star$ & $90.384$   \\ \hline
        6 & $\Delta_6^\star$ & $111.175$   \\ \hline
        \hline 
\end{tabular}
\caption{Table \ref{table:dce} completed with the DCE of higher $n$ resonances of the $\Delta$ baryon family.} 
\label{table:extdce}
\end{center}
\end{table}

\begin{figure}[H]
	\centering
	\includegraphics[width=10cm]{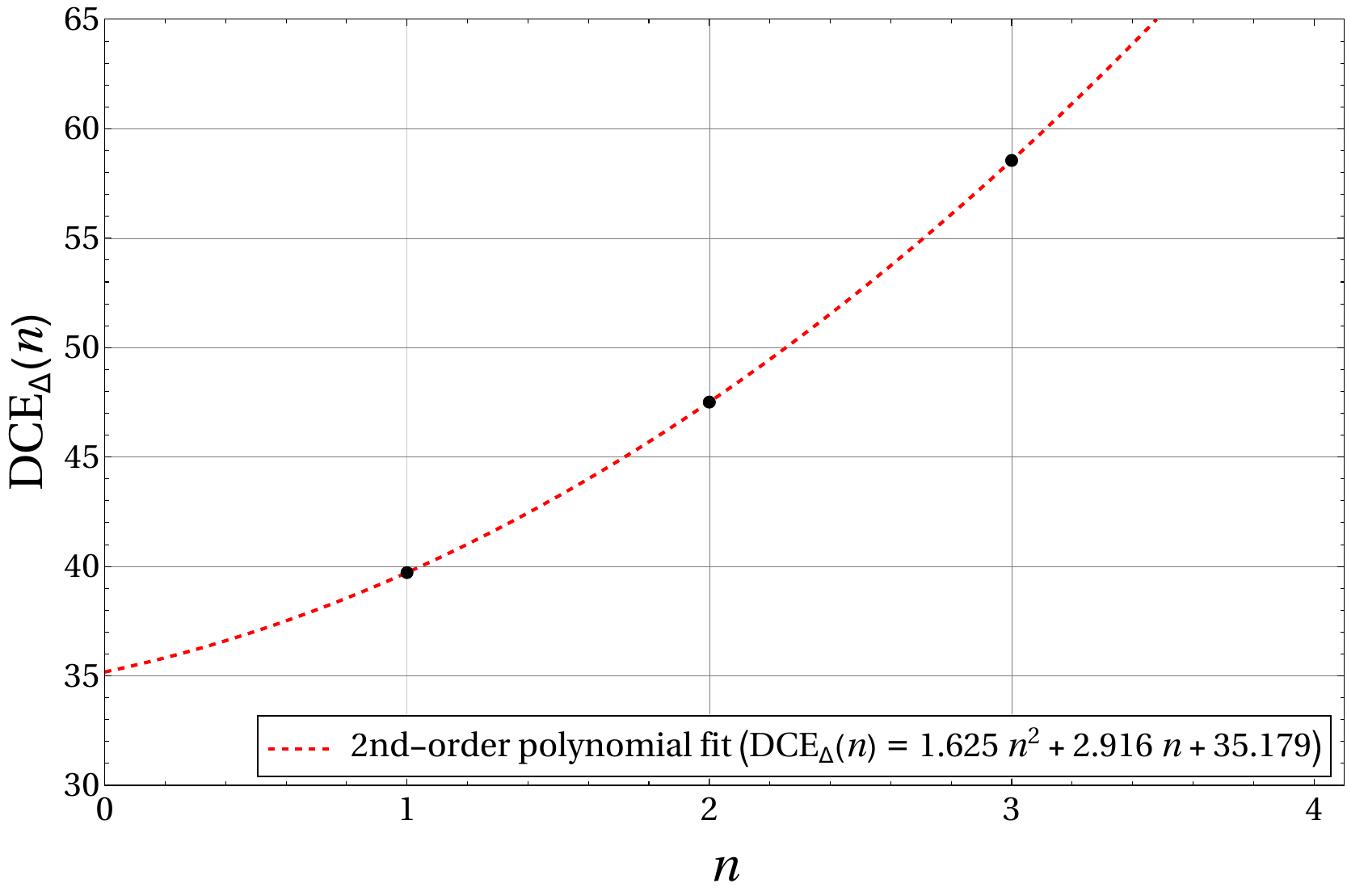}
	\caption{DCE of the $\Delta$ baryon resonances as a function of the radial quantum number.}
	\label{fig:dcen}
\end{figure}

Complementarily, the second type of DCE–Regge-like trajectory interpolates the DCE values for the same set of resonances, but this time as a function of the mass.
This interpolation, shown in Fig.~\ref{fig:dcem}, is described by the following equation:
\begin{flalign}
    \text{DCE}_\Delta (m^2) =  0.3747 m^4 + 6.7341 m^2 + 28.6356\,.
    \label{dcem}
\end{flalign}
also with a RMSD within $10^{-14}$ \%.

\begin{figure}[H]
	\centering
	\includegraphics[width=10cm]{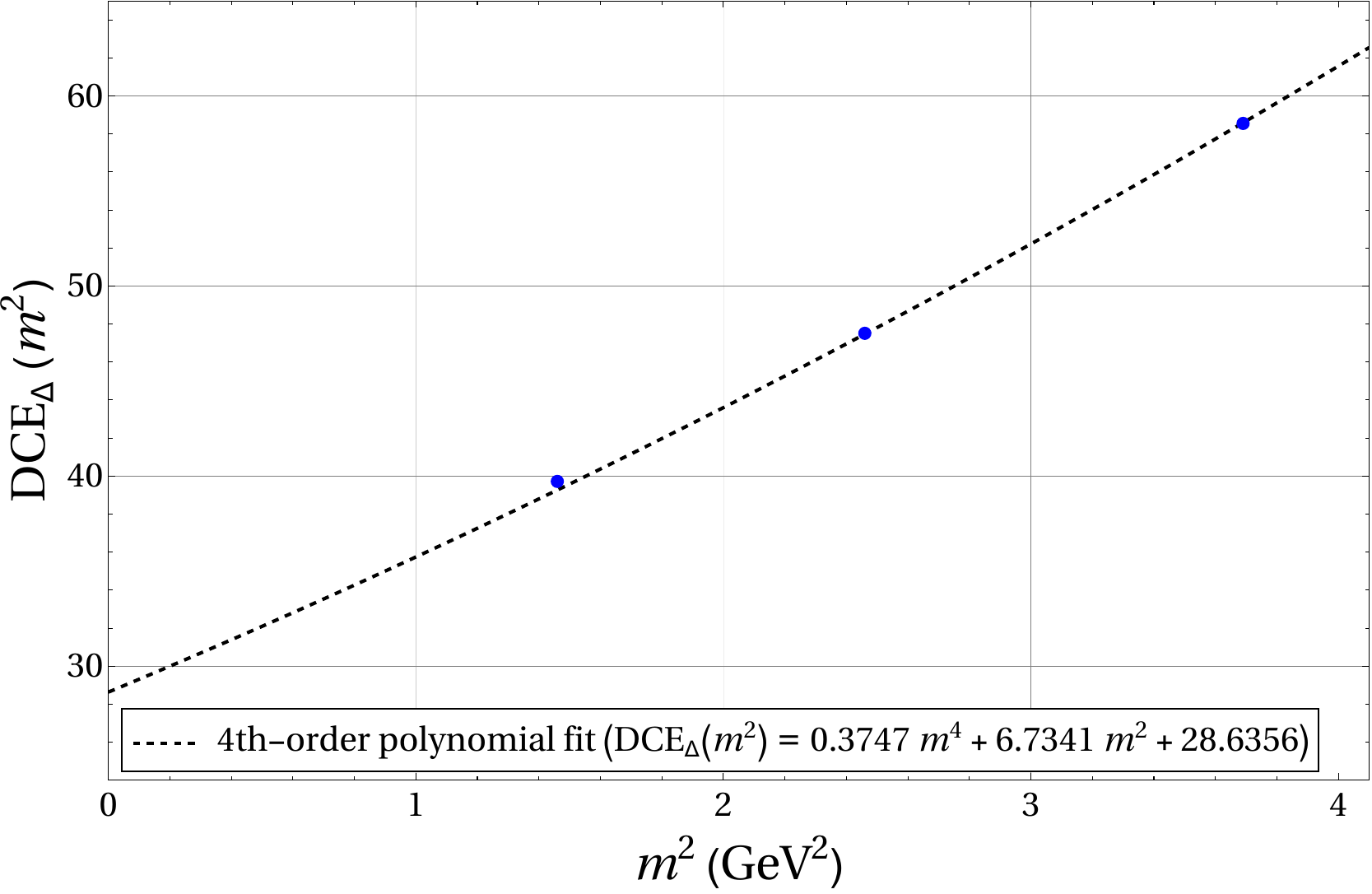}
	\caption{DCE of the $\Delta$ baryon resonances as a function of their squared mass, for $n=1,2,3$.}
	\label{fig:dcem}
\end{figure}

The next step, based on the two types of trajectories obtained, is to determine the extrapolated masses of the higher excited states. The DCE corresponding to the first extrapolated state, $\Delta_4^\star$, is found by inserting $n=4$ into Eq.~(\ref{dcen}), and similarly for $n=5$ and $n=6$, as presented in Table \ref{table:extdce}. Then, for each state, the calculated DCE value is substituted into the left-hand side of Eq.~(\ref{dcem}), resulting in a quadratic algebraic equation. Solving this equation for $m$ yields two solutions: one is physically inconsistent and can be discarded, while the other corresponds to the estimated mass of the extrapolated state.

By inserting $n=4$ into Eq.~(\ref{dcen}), one obtains a DCE value of $72.843$ nat, which, when substituted into Eq.~(\ref{dcem}), results in an estimated mass for the $\Delta_4^\star$ state of $2261 \pm 32$ MeV. For $n=5$, the same procedure yields a DCE value of $90.384$ nat, leading to an estimated mass for the $\Delta_5^\star$ state of $2585 \pm 36$ MeV. Finally, for $n=6$, the DCE value is $111.175$ nat, which results in an estimated mass for the $\Delta_6^\star$ state of $2892 \pm 40$ MeV. 
Since the Regge-like trajectory equations were built by fitting experimental mass data, the DCE-based AdS/QCD hybrid approach is firmly anchored in observed results. This suggests that the mass predictions produced by this method, shown in Table~\ref{table:dcemass}, offer a more accurate reflection of the physical spectrum than estimates obtained by directly applying Eq.~(\ref{schrodingerlike}) to determine masses from the quantum number $n$.

\begin{table}[H]
\begin{center}
\begin{tabular}{||c|c|c|c||}
\hline\hline
        $\quad n \quad$ &\; State \;&\; $M_\text{exp}$ (MeV) \;&\; $M_\text{AdS/QCD}$ (MeV)\; \\\hline\hline \hline
        1 &\; $\Delta(1232)$ \;& \textcolor{black}{$1210 \pm 10$} & $900 \pm 1$ \\ \hline
        2 & $\Delta(1600)$ & $1570 \pm 70$ & $1390 \pm 26$ \\ \hline
        3 & $\Delta(1920)$ & $1920 \pm 50$ & $1670 \pm 41$ \\ \hline 
        4 & $\Delta_4^\star$ & --- & $2261 \pm 32^\star$ \\ \hline
        5 & $\Delta_5^\star$ & --- & $2585 \pm 36^\star$ \\ \hline
        6 & $\Delta_6^\star$ & --- & $2892 \pm 40^\star$ \\ \hline
        \hline 
\end{tabular}
\caption{ Table~\ref{table:mass} completed with the higher $n$ resonances of the $\Delta$ baryon family. The extrapolated masses for $n = 4,5,6$, indicated with the $\star$, were obtained through the combined use of Eqs. (\ref{dcen}, \ref{dcem}).} 
\label{table:dcemass}
\end{center}
\end{table}

Ideally, the extrapolated mass values obtained here are compared with the experimentally detected masses of further states listed in the PDG, which have not yet been fully identified.
However, higher baryonic states are scarcer than, for instance, mesonic ones, resulting in a much smaller set of available unidentified candidates.
When the search is further constrained by matching the specific quantum numbers and quark composition, the number of possible options decreases even more, although it remains feasible to find plausible correspondences.
The estimated mass for the $\Delta_6^\star$ state, $2892 \pm 40$ MeV, lies within the uncertainty range of higher-mass candidates for spin-$3/2$ $\Delta$ baryons, grouped in the PDG as $\Delta(\sim 3000)$, with $m = 2850 \pm 150$ MeV. This result may provide future insights into states in this high-energy region.

\section{DCC of Spin-$3/2$ $\Delta$ baryon resonances in AdS/QCD}\label{section:dcc}
\label{sec21}
Complementary to the DCE, Ref.~\cite{Gleiser:2018jpd} introduced an additional
CIM, namely the DCC.
Its definition is based on a modal fraction whose normalization differs from that used in the DCE.
Following the same initial steps as in constructing the DCE, one first evaluates the Fourier transform in Eq.~(\ref{fourier}). The modal fraction relevant for the DCC is then defined as
\begin{equation}\label{mf2}
\mathsf{f}_{00}(k_A)
=
\frac{\left|\langle \tau_{00}(k_A)\rangle\right|^{2}}
{\left|\langle{\tau_{00}^{\textsc{max}}}(k_A)\rangle\right|^{2}}\,,
\end{equation}
where $\langle{\tau_{00}^{\textsc{max}}}(k_A)\rangle$ represents the largest value of $\left|\langle \tau_{00}(k_A)\rangle\right|$ in momentum space.
The modal fraction~(\ref{mf2}) is useful to measure the contribution of each wave mode to the power spectral density of the energy density.
Configurations with uniform spectral distributions exhibit lower complexity, while spectra that are more uneven correspond to higher intrinsic complexity.

The DCC is then defined as the Shannon entropy associated with $\mathsf{f}_{00}(k_A)$~\cite{Gleiser:2018jpd} as
\begin{equation}\label{dcc}
S_{\textsc{DCC}}
= -\int_{\rm AdS}
\mathsf{f}_{00}(k_A)
\log \mathsf{f}_{00}(k_A)\,
\sqrt{g}\,dk_A\,.
\end{equation}

The DCC equals zero when each nonzero mode contributes equally to the spectrum.
Thus, a configuration with a uniform power spectral density has maximal DCE but a vanishing DCC, whereas a single plane-wave setup yields zero values for both the DCE and DCC. \clt{
The distinction between DCE and DCC arises from the different normalization conditions of their modal fractions. While the DCE modal fraction satisfies a fixed normalization [see Eq.~(\ref{dce_modal})], the DCC modal fraction in Eq.~(\ref{mf2}) is normalized by a functional quantity, encoding the maximal spectral contribution for the given configuration. This ensures the DCC is always non-negative, in contrast to the DCE, and implies that the DCC only vanishes when all modes contribute equally~\cite{Gleiser:2018jpd}.} The DCC can be interpreted as a measure of configurational complexity associated with shape. The shape of a physical configuration is encoded in its two-point correlator, whose Fourier transform gives the power spectral density. Consequently, the DCC quantifies how the geometrical features of a configuration are distributed across momentum modes~\cite{Gleiser:2018jpd}.
\begin{table}[H]
\begin{center}
\begin{tabular}{||c|c|c||}
\hline\hline
        $\quad n \quad$ &\; State \;&\; DCC (nat) \;\\\hline\hline \hline
        1 &\; $\Delta(1232)$ \;& $68.791$ \\ \hline
        2 & $\Delta(1600)$ & $80.433$ \\ \hline
        3 & $\Delta(1920)$ & $95.659$   \\ \hline\hline 
\end{tabular}
\caption{DCC of the radial resonances of the $\Delta$ baryon family.} 
\label{table:dcc}
\end{center}
\end{table}

The analysis from this point onward is analogous to that performed for the DCE.
From the DCC values, it is possible to construct the first type of DCC-Regge-like trajectory, which describes the DCC as a function of the radial quantum number $n$.
This expression is obtained by interpolating the DCC values, as shown in Fig.~\ref{fig:dccn}, and is given by
\begin{flalign}
    \text{DCC}_\Delta (n) = 1.792 n^2 + 6.266 n + 60.733\,,
    \label{dccn}
\end{flalign}
with a RMSD within $10^{-14}$ \%.
With this relation, the DCC values for the states $n=4,5,6$ can be calculated, as presented in Table~\ref{table:extdcc}.

\begin{table}[H]
\begin{center}
\begin{tabular}{||c|c|c||}
\hline\hline
        $\quad n \quad$ &\; State \;&\; DCC (nat) \;\\\hline\hline \hline
        1 &\; $\Delta(1232)$ \;& $68.791$ \\ \hline
        2 & $\Delta(1600)$ & $80.433$ \\ \hline
        3 & $\Delta(1920)$ & $95.659$   \\ \hline
        4 & $\Delta_4^\star$ & $114.469$   \\ \hline
        5 & $\Delta_5^\star$ & $136.863$   \\ \hline
        6 & $\Delta_6^\star$ & $162.841$   \\ \hline
        \hline 
\end{tabular}
\caption{Table~\ref{table:dcc} completed with the DCC of higher $n$ resonances of the $\Delta$ baryon family.} 
\label{table:extdcc}
\end{center}
\end{table}

\begin{figure}[H]
	\centering
	\includegraphics[width=10cm]{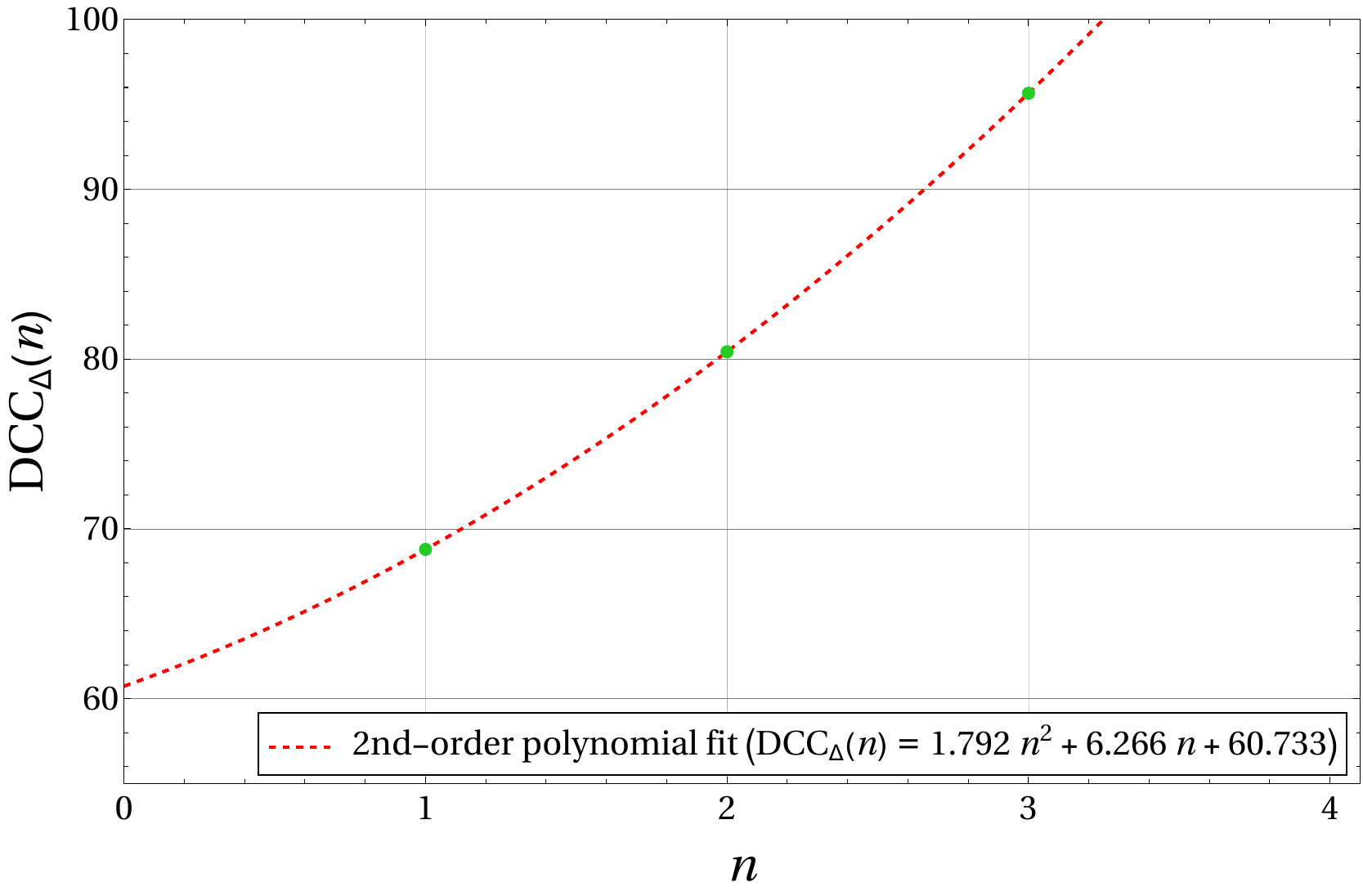}
	\caption{DCC of the $\Delta$ baryon resonances as a function of the radial quantum number.}
	\label{fig:dccn}
\end{figure}

The second type of DCC-Regge-like trajectory is computed in the same way, but interpolating the DCC values with respect to the squared mass of the resonances.
Fig.~\ref{fig:dccm} shows this interpolation, and the resulting equation is written as
\begin{flalign}
    \text{DCC}_\Delta (m^2) = 0.0795 m^4 + 11.9759 m^2 + 50.4305\,.
    \label{dccm}
\end{flalign}
with a root-mean-square deviation (RMSD) within $10^{-14}$ \%.

\begin{figure}[H]
	\centering
	\includegraphics[width=10cm]{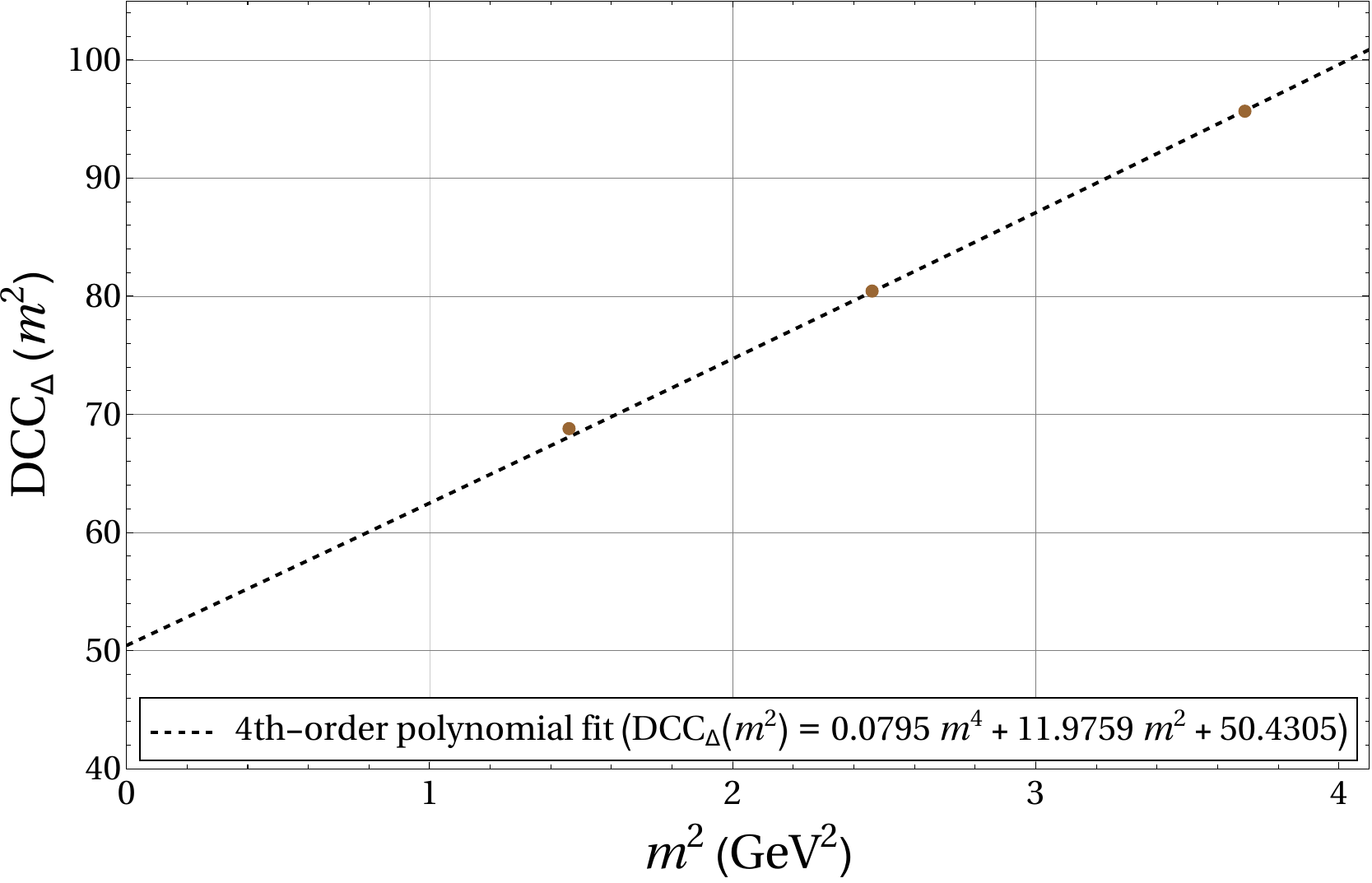}
	\caption{DCC of the $\Delta$ baryon resonances as a function of their squared mass, for $n=1,2,3$.}
	\label{fig:dccm}
\end{figure}

Using Eqs.~(\ref{dccn}, \ref{dccm}), we repeat the same procedure as in the previous section, which consists of taking the DCC value obtained from Eq.~(\ref{dccn}) for a given extrapolated state, inserting it into Eq.~(\ref{dccn}), and solving it to find the estimated mass of that state.
The estimated masses resulting from this process are summarized in Table~\ref{table:dccmass}.

\begin{table}[H]
\begin{center}
\begin{tabular}{||c|c|c|c||}
\hline\hline
        $\quad n \quad$ &\; State \;&\; $M_\text{exp}$ (MeV) \;&\; $M_\text{AdS/QCD}$ (MeV)\; \\\hline\hline \hline
        1 &\; $\Delta(1232)$ \;& \textcolor{black}{$1210 \pm 10$} & $900 \pm 1$ \\ \hline
        2 & $\Delta(1600)$ & $1570 \pm 70$ & $1390 \pm 26$ \\ \hline
        3 & $\Delta(1920)$ & $1920 \pm 50$ & $1670 \pm 41$ \\ \hline\
        4 & $\Delta_4^\star$ & --- & $2273 \pm 32^\star$ \\ \hline
        5 & $\Delta_5^\star$ & --- & $2627 \pm 37^\star$ \\ \hline
        6 & $\Delta_6^\star$ & --- & $2977 \pm 42^\star$ \\ \hline
        \hline 
\end{tabular}
\caption{ Table~\ref{table:mass} completed with the higher $n$ resonances of the $\Delta$ baryon family. The extrapolated masses for $n = 4,5,6$, indicated with the $\star$, were obtained through the combined use of Eqs.~(\ref{dccn},\ref{dccm}).} 
\label{table:dccmass}
\end{center}
\end{table}

The complementary analysis using the DCC reinforces the result previously obtained with the DCE, suggesting a possible correspondence between the $\Delta_6^\star$ state and the $\Delta(\sim3000)$ resonance, as indicated by the overlapping mass range, $2977 \pm 42$ MeV and $2850 \pm 150$ MeV, respectively.
This reinforcement highlights the importance of the combined use of DCC in the analysis and contributes to a more precise direction for future insights. 

\section{Conclusions}\label{section:conclusion}
In this work, we explored the mass spectrum and the informational content of spin-$3/2$ $\Delta$ baryon resonances using a combined approach based on AdS/QCD and CIMs, namely the DCE and the DCC. The linear Regge trajectory, Eq.~(\ref{schrodingerlike}), successfully interpolates the experimental squared masses as a function of the radial quantum number $n$, while the AdS/QCD model provides a holographic description of the $\Delta$ baryons via Rarita--Schwinger fields. The resulting energy density along the holographic direction, Eq.~(\ref{t000}), encodes both the spin and radial structure of the baryons, linking bulk dynamics to boundary phenomenology.

The DCE, computed from the energy density of the Rarita--Schwinger fields along the AdS bulk, provides a quantitative measure of the spatial informational content carried by each baryonic state. By analyzing the three experimentally established $\Delta$ baryon resonances, we constructed two types of Regge-like trajectories: one as a function of the radial excitation number, Eq.~(\ref{dcen}), and another as a function of the squared mass, Eq.~(\ref{dcem}). These trajectories enabled a reliable extrapolation to heavier $\Delta$ baryon states, yielding mass estimates for the next radial excitations, $\Delta_4^\star$, $\Delta_5^\star$, and $\Delta_6^\star$, summarized in Table~\ref{table:dcemass}.

Complementarily, the DCC was analyzed for the same set of states. Unlike the DCE, the DCC employs a modal fraction normalized to the maximal spectral contribution, ensuring positivity and emphasizing how spectral weight is distributed among momentum modes. While the DCE quantifies the overall spatial informational content, the DCC characterizes the configurational complexity of shape, highlighting the geometric distribution of energy in momentum space. Using the DCC, analogous Regge-like trajectories were constructed as functions of both the radial quantum number [Eq.~(\ref{dccn})] and the squared mass [Eq.~(\ref{dccm})], allowing an independent estimation of higher radial excitations. In particular, the DCC-based extrapolation predicts a mass for the $\Delta_6^\star$ state of $2977 \pm 42$ MeV, consistent with high-mass candidates around $2850 \pm 150$ MeV reported by the PDG.

Several key aspects emerge from this analysis. First, the DCE and DCC provide complementary insights into the physical relevance of baryonic states. Lower DCE values correlate with more stable, experimentally accessible resonances, while the DCC captures the nonuniformity of momentum-mode distributions and thus probes internal shape complexity of $\Delta$ baryon resonances. Second, the holographic description highlights the interplay between AdS curvature, spin-$3/2$ constraints, and the two independent bulk mass parameters $(m_1,m_2)$, which together determine the radial profiles and energy distributions of the $\Delta$ baryons. Third, the Regge-like trajectories furnish a systematic framework for predicting heavier excitations in the $\Delta$ baryonic spectra. Both the DCE and DCC can serve as effective tools in hadron spectroscopy, complementing traditional mass-based analyses. The methodology can be extended to other baryonic families, including hyperons and heavier quark sectors, and may offer insights into the role of excited baryons in dense environments, such as neutron-star matter, where $\Delta$ resonances are known to influence the equation of state and compact-star properties.

From a theoretical standpoint, the analysis clarifies how the microscopic holographic construction underlies the emergent informational patterns observed in the $\Delta$ baryon spectrum. The combined effects of the hard-wall geometry, the scalar background $X_0(z)$ responsible for chiral symmetry breaking, and the Rarita--Schwinger dynamics uniquely fix the radial profiles of spin-$3/2$ modes and their associated energy densities. The presence of two independent bulk mass parameters $(m_1,m_2)$, together with the IR rescaling parameter $\xi$, encodes the distinction between mesonic and baryonic confinement scales within a unified AdS/QCD framework.
The consistency between holographic dynamics and CIMs suggests that the DCE and DCC provide sensitive probes of how bulk geometric and field-theoretic structures map onto physically relevant boundary spectra. The framework developed here can be systematically extended to other higher-spin baryons and to alternative holographic realizations, including soft-wall, finite-density and magnetized backgrounds, such as the self-consistent constructions of Refs.~\cite{Toniato:2025gts,Jena:2024cqs}, where modifications of the IR geometry and scalar condensates are expected to leave characteristic imprints on configurational observables.

From a phenomenological perspective, the extrapolated masses for the higher radial excitations $\Delta_4^\star$, $\Delta_5^\star$, and $\Delta_6^\star$ populate an energy region where experimental information is limited, and resonances are often poorly established. The identification of highly excited spin-$3/2$ $\Delta$ baryons remains experimentally challenging due to their large decay widths and strong overlap with neighboring states. In this context, the information-theoretic observables employed here offer a complementary perspective: the monotonic growth of both the DCE and DCC with the radial quantum number reflects the increasing spatial extension and configurational complexity of higher excitations, which may be indirectly manifested in experimental observables such as broader decay patterns and enhanced couplings to multi-hadron channels.

The combined AdS/QCD and configurational-information framework developed in this work provides a unified and experimentally anchored approach to the spectroscopy of spin-$3/2$ $\Delta$ baryons. Its predictive capability for higher excited states and its sensitivity to the internal structure of baryonic configurations make it a promising tool for guiding future experimental searches and refining the phenomenological understanding of $\Delta$ baryon resonances in the high-energy regime.

As a perspective, near the critical temperature $T_c$, thermal effects are expected to modify the holographic energy density profiles of spin-$3/2$ $\Delta$ baryons systematically. As $T$ approaches $T_c$ from below, the weakening of confinement, the partial melting of the chiral condensate, and the emergence of a black hole horizon in finite-temperature AdS backgrounds tend to suppress the IR support of the $\Delta$ baryonic profiles, leading to broader and less localized energy density distributions along the holographic direction. These effects are expected to be more pronounced for higher radial excitations, whose profiles extend deeper into the IR regime. From an information-theoretic perspective, such thermal modifications would correspond to a reduction or saturation of the DCE, accompanied by a redistribution of momentum-space modes that may enhance the DCC. Together, these features can provide a qualitative signature of the gradual delocalization and eventual dissociation of $\Delta$ baryon resonances as the system approaches the deconfinement transition, as already implemented for nucleonic baryons in Refs. \cite{Ferreira:2020iry,daRocha:2025gcz}. 
These studies analyzed the nucleonic baryon resonances mass spectra as a function of the hot medium temperature, and also revealed similar stability at low temperatures. It was also observed a baryonic phase transition slightly above the Hagedorn
temperature, supporting that the DCE
accurately measures the stability of baryons during a phase transition.
A quantitative assessment of these effects for the $\Delta$ baryons requires explicit finite-temperature holographic constructions, which are left for future investigations.

\subsection*{Acknowledgments}
HA is supported by the São Paulo Research Foundation (FAPESP) 
(Grant No. 2025/01268-4). 
RdR~thanks to FAPESP (Grants No. 2021/01089-1, No. 2025/23004-9, and No.~2024/05676-7), and to the National Council for Scientific and Technological Development (CNPq) (Grants No. 303742/2023-2 and No. 401567/2023-0), for partial financial support. PHOS and BT both thank CAPES - Brazil - Finance Code 001 and UFABC.

\bibliography{bibliography}

@article{PDG2024,
  author       = {Particle Data Group and others},
  title        = {Review of Particle Physics},
  journal      = {Prog. Theor. Exp. Phys.},
  year         = {2024},
  volume       = {2024},
  number       = {8},
  pages        = {083C01},
  doi          = {10.1093/ptep/ptae083},
}

@article{Andronic2018,
  author       = {A. Andronic and P. Braun-Munzinger and K. Redlich and J. Stachel},
  title        = {Decoding the phase structure of QCD via particle production at high energy},
  journal      = {Nature},
  year         = {2018},
  volume       = {561},
  pages        = {321--330},
  doi          = {10.1038/s41586-018-0491-6},
}

@article{Jena:2024cqs,
    author = "Jena, Siddhi Swarupa and Barman, Jyotirmoy and Toniato, Bruno and Dudal, David and Mahapatra, Subhash",
    title = "{A dynamical Einstein-Born-Infeld-dilaton model and holographic quarkonium melting in a magnetic field}",
    eprint = "2408.14813",
    archivePrefix = "arXiv",
    primaryClass = "hep-th",
    doi = "10.1007/JHEP12(2024)096",
    journal = "JHEP",
    volume = "12",
    pages = "096",
    year = "2024"
}

@article{Dudal:2014jfa,
    author = "Dudal, David and Mertens, Thomas G.",
    title = "{Melting of charmonium in a magnetic field from an effective AdS/QCD model}",
    eprint = "1410.3297",
    archivePrefix = "arXiv",
    primaryClass = "hep-th",
    doi = "10.1103/PhysRevD.91.086002",
    journal = "Phys. Rev. D",
    volume = "91",
    pages = "086002",
    year = "2015"
}

@article{Dudal:2018rki,
    author = "Dudal, David and Mertens, Thomas G.",
    title = "{Holographic estimate of heavy quark diffusion in a magnetic field}",
    eprint = "1802.02805",
    archivePrefix = "arXiv",
    primaryClass = "hep-th",
    doi = "10.1103/PhysRevD.97.054035",
    journal = "Phys. Rev. D",
    volume = "97",
    number = "5",
    pages = "054035",
    year = "2018"
}

@article{Boschi-Filho:2002xih,
    author = "Boschi-Filho, Henrique and Braga, Nelson R. F.",
    title = "{Gauge / string duality and scalar glueball mass ratios}",
    eprint = "hep-th/0212207",
    archivePrefix = "arXiv",
    doi = "10.1088/1126-6708/2003/05/009",
    journal = "JHEP",
    volume = "05",
    pages = "009",
    year = "2003"
}

@article{dePaula:2009za,
    author = "de Paula, W. and Frederico, T.",
    title = "{Scalar mesons within a dynamical holographic QCD model}",
    eprint = "0908.4282",
    archivePrefix = "arXiv",
    primaryClass = "hep-ph",
    doi = "10.1016/j.physletb.2010.08.045",
    journal = "Phys. Lett. B",
    volume = "693",
    pages = "287--291",
    year = "2010"
}

@article{dePaula:2008fp,
    author = "de Paula, W. and Frederico, T. and Forkel, H. and Beyer, M.",
    title = "{Dynamical AdS/QCD with area-law confinement and linear Regge trajectories}",
    eprint = "0806.3830",
    archivePrefix = "arXiv",
    primaryClass = "hep-ph",
    doi = "10.1103/PhysRevD.79.075019",
    journal = "Phys. Rev. D",
    volume = "79",
    pages = "075019",
    year = "2009"
}

@article{Braga:2018hjt,
    author = "Braga, Nelson R. F. and Ferreira, Luiz F.",
    title = "{Quasinormal modes and dispersion relations for quarkonium in a plasma}",
    eprint = "1810.11872",
    archivePrefix = "arXiv",
    primaryClass = "hep-ph",
    doi = "10.1007/JHEP01(2019)082",
    journal = "JHEP",
    volume = "01",
    pages = "082",
    year = "2019"
}

@article{Braga:2023qee,
    author = "Braga, Nelson R. F. and Ferreira, Luiz F. and Junqueira, Octavio C.",
    title = "{Configuration entropy of a rotating quark-gluon plasma from holography}",
    eprint = "2301.01322",
    archivePrefix = "arXiv",
    primaryClass = "hep-th",
    doi = "10.1016/j.physletb.2023.138265",
    journal = "Phys. Lett. B",
    volume = "847",
    pages = "138265",
    year = "2023"
}

@article{BallonBayona:2007vp,
    author = "Ballon Bayona, C. A. and Boschi-Filho, Henrique and Braga, Nelson R. F. and Pando Zayas, Leopoldo A.",
    title = "{On a Holographic Model for Confinement/Deconfinement}",
    eprint = "0705.1529",
    archivePrefix = "arXiv",
    primaryClass = "hep-th",
    reportNumber = "MCTP-07-17",
    doi = "10.1103/PhysRevD.77.046002",
    journal = "Phys. Rev. D",
    volume = "77",
    pages = "046002",
    year = "2008"
}

@article{Ballon-Bayona:2017bwk,
    author = "Ballon-Bayona, Alfonso and Krein, Gastao and Miller, Carlisson",
    title = "{Strong couplings and form factors of charmed mesons in holographic QCD}",
    eprint = "1702.08417",
    archivePrefix = "arXiv",
    primaryClass = "hep-ph",
    doi = "10.1103/PhysRevD.96.014017",
    journal = "Phys. Rev. D",
    volume = "96",
    number = "1",
    pages = "014017",
    year = "2017"
}

@article{Parmar:2025csx,
    author = "Parmar, Vishal and Thapa, Vivek Baruah and Sinha, Monika and Bombaci, Ignazio",
    title = "{Exploring the {\ensuremath{\Delta}}-resonance in neutron stars: Implications from astrophysical and nuclear observations}",
    eprint = "2503.07256",
    archivePrefix = "arXiv",
    primaryClass = "astro-ph.HE",
    doi = "10.1103/643w-c2ly",
    journal = "Phys. Rev. D",
    volume = "112",
    number = "2",
    pages = "023016",
    year = "2025"
}

@article{Rougemont:2015ona,
    author = "Rougemont, Romulo and Noronha, Jorge and Noronha-Hostler, Jacquelyn",
    title = "{Suppression of baryon diffusion and transport in a baryon rich strongly coupled quark-gluon plasma}",
    eprint = "1507.06972",
    archivePrefix = "arXiv",
    primaryClass = "hep-ph",
    doi = "10.1103/PhysRevLett.115.202301",
    journal = "Phys. Rev. Lett.",
    volume = "115",
    number = "20",
    pages = "202301",
    year = "2015"
}

@article{Capstick:1986bm,
  author = {Capstick, Simon and Isgur, Nathan},
  title = {Baryons in a Relativized Quark Model with Chromodynamics},
  journal = {Phys. Rev. D},
  volume = {34},
  pages = {2809--2835},
  year = {1986}
}

@article{Henneaux:1985kr,
  author    = {M. Henneaux and C. Teitelboim},
  title     = {Dynamics of the gravitational field in the Hamiltonian formalism},
  journal   = {Comm. Math.  Phys.},
  volume    = {98},
  pages     = {391--424},
  year      = {1985}
}

@article{Gentile:2012jm,
    author = "Gentile, L. G. C. and Grassi, P. A. and Mezzalira, A.",
    title = "{Fermionic Wigs for AdS-Schwarzschild Black Holes}",
    eprint = "1207.0686",
    archivePrefix = "arXiv",
    primaryClass = "hep-th",
    reportNumber = "DFTT-8-2012, DISIT-2012, DFPD-12-TH-6",
    doi = "10.1007/JHEP10(2013)065",
    journal = "JHEP",
    volume = "10",
    pages = "065",
    year = "2013"
}

@article{Shaukat:2009hp,
    author = "Shaukat, Abrar and Waldron, Andrew",
    title = "{Weyl's Gauge Invariance: Conformal Geometry, Spinors, Supersymmetry, and Interactions}",
    eprint = "0911.2477",
    archivePrefix = "arXiv",
    primaryClass = "hep-th",
    doi = "10.1016/j.nuclphysb.2009.11.020",
    journal = "Nucl. Phys. B",
    volume = "829",
    pages = "28--47",
    year = "2010"
}

@article{Matlock:1999fy,
    author = "Matlock, P. and Viswanathan, K. S.",
    title = "{The AdS / CFT correspondence for the massive Rarita-Schwinger field}",
    eprint = "hep-th/9906077",
    archivePrefix = "arXiv",
    doi = "10.1103/PhysRevD.61.026002",
    journal = "Phys. Rev. D",
    volume = "61",
    pages = "026002",
    year = "2000"
}

@article{Koshelev:1998tu,
    author = "Koshelev, A. S. and Rytchkov, O. A.",
    title = "{Note on the massive Rarita-Schwinger field in the AdS / CFT correspondence}",
    eprint = "hep-th/9812238",
    archivePrefix = "arXiv",
    reportNumber = "SMI-25-98",
    doi = "10.1016/S0370-2693(99)00148-3",
    journal = "Phys. Lett. B",
    volume = "450",
    pages = "368--376",
    year = "1999"
}

@article{Volovich:1998tj,
    author = "Volovich, Anastasia",
    title = "{Rarita-Schwinger field in the AdS / CFT correspondence}",
    eprint = "hep-th/9809009",
    archivePrefix = "arXiv",
    doi = "10.1088/1126-6708/1998/09/022",
    journal = "JHEP",
    volume = "09",
    pages = "022",
    year = "1998"
}

@article{Gomes:2023qkj,
    author = "Gomes, M. and Lima, J. G. and Mariz, T. and Nascimento, J. R. and Petrov, A. Yu.",
    title = "{Non-Abelian Carroll{\textendash}Field{\textendash}Jackiw term in a Rarita-Schwinger model}",
    eprint = "2308.16308",
    archivePrefix = "arXiv",
    primaryClass = "hep-th",
    doi = "10.1016/j.physletb.2023.138141",
    journal = "Phys. Lett. B",
    volume = "845",
    pages = "138141",
    year = "2023"
}

@article{Dantas:2015dca,
    author = "Dantas, D. M. and Veras, D. F. S. and Silva, J. E. G. and Almeida, C. A. S.",
    title = "{Fermionic Kaluza-Klein modes in the string-cigar braneworld}",
    eprint = "1506.07228",
    archivePrefix = "arXiv",
    primaryClass = "hep-th",
    doi = "10.1103/PhysRevD.92.104007",
    journal = "Phys. Rev. D",
    volume = "92",
    number = "10",
    pages = "104007",
    year = "2015"
}

@article{Kirsch:2006he,
    author = "Kirsch, Ingo",
    title = "{Spectroscopy of fermionic operators in AdS/CFT}",
    eprint = "hep-th/0607205",
    archivePrefix = "arXiv",
    reportNumber = "HUTP-06-A0028",
    doi = "10.1088/1126-6708/2006/09/052",
    journal = "JHEP",
    volume = "09",
    pages = "052",
    year = "2006"
}

@article{Gursoy:2007cb,
    author = "Gursoy, U. and Kiritsis, E.",
    title = "{Exploring improved holographic theories for QCD: Part I}",
    eprint = "0707.1324",
    archivePrefix = "arXiv",
    primaryClass = "hep-th",
    reportNumber = "CPHT-RR027-0507",
    doi = "10.1088/1126-6708/2008/02/032",
    journal = "JHEP",
    volume = "02",
    pages = "032",
    year = "2008"
}

@article{Meert:2018qzk,
    author = "Meert, P. and da Rocha, R.",
    title = "{The emergence of flagpole and flag-dipole fermions in fluid/gravity correspondence}",
    eprint = "1809.01104",
    archivePrefix = "arXiv",
    primaryClass = "hep-th",
    doi = "10.1140/epjc/s10052-018-6497-2",
    journal = "Eur. Phys. J. C",
    volume = "78",
    number = "12",
    pages = "1012",
    year = "2018"
}

@article{CoronadoVillalobos:2015mns,
    author = "Coronado Villalobos, C. H. and Hoff da Silva, J. M. and da Rocha, Rold{\~a}o",
    title = "{Questing mass dimension 1 spinor fields}",
    eprint = "1504.06763",
    archivePrefix = "arXiv",
    primaryClass = "hep-th",
    doi = "10.1140/epjc/s10052-015-3498-2",
    journal = "Eur. Phys. J. C",
    volume = "75",
    number = "6",
    pages = "266",
    year = "2015"
}

@article{Karch:2006pv,
  author = {Karch, Andreas and Katz, Emanuel and Son, Dam T. and Stephanov, Mikhail A.},
  title = {Linear Confinement and AdS/QCD},
  journal = {Phys. Rev. D},
  volume = {74},
  pages = {015005},
  year = {2006},
  eprint = {hep-ph/0602229}
}

@article{Gleiser:2011di,
  author = {Gleiser, Marcelo and Sowinski, Nicholas},
  title = {Information-Entropic Signature of the Critical Point},
  journal = {Phys. Lett. B},
  volume = {747},
  pages = {125--128},
  year = {2015},
  eprint = {1501.06800}
}

@article{Gleiser:2012tu,
  author = {Gleiser, Marcelo and Stamatopoulos, Nan},
  title = {Information Content of Spontaneous Symmetry Breaking},
  journal = {Phys. Rev. D},
  volume = {86},
  pages = {045004},
  year = {2012},
  eprint = {1205.3061}
}

@article{MartinContreras:2023oqs,
    author = "Martin Contreras, Miguel Angel and Vega, Alfredo",
    title = "{Holographic stability for non-qq\textasciimacron{} candidates}",
    eprint = "2309.02905",
    archivePrefix = "arXiv",
    primaryClass = "hep-ph",
    doi = "10.1103/PhysRevD.108.126024",
    journal = "Phys. Rev. D",
    volume = "108",
    number = "12",
    pages = "126024",
    year = "2023"
}

@article{Gleiser:2018jpd,
    author = "Gleiser, Marcelo and Sowinski, Damian",
    title = "{Configurational information approach to instantons and false vacuum decay in $D$-dimensional spacetime}",
    eprint = "1807.07588",
    archivePrefix = "arXiv",
    primaryClass = "hep-th",
    doi = "10.1103/PhysRevD.98.056026",
    journal = "Phys. Rev. D",
    volume = "98",
    number = "5",
    pages = "056026",
    year = "2018"
}

@article{Marquez:2022gmu,
    author = "Marquez, Kauan D. and Menezes, D{\'e}bora P. and Pais, Helena and Provid{\^e}ncia, Constan{\c{c}}a",
    title = "{{\ensuremath{\Delta}} baryons in neutron stars}",
    eprint = "2206.02935",
    archivePrefix = "arXiv",
    primaryClass = "nucl-th",
    doi = "10.1103/PhysRevC.106.055801",
    journal = "Phys. Rev. C",
    volume = "106",
    number = "5",
    pages = "055801",
    year = "2022"
}

@article{Bernardini:2018uuy,
    author = "Bernardini, A. E. and da Rocha, Roldao",
    title = "{Informational entropic Regge trajectories of meson families in AdS/QCD}",
    eprint = "1809.10055",
    archivePrefix = "arXiv",
    primaryClass = "hep-th",
    doi = "10.1103/PhysRevD.98.126011",
    journal = "Phys. Rev. D",
    volume = "98",
    number = "12",
    pages = "126011",
    year = "2018"
}

@article{Pirner:2009gr,
    author = "Pirner, H. J. and Galow, B.",
    title = "{Strong Equivalence of the AdS-Metric and the QCD Running Coupling}",
    eprint = "0903.2701",
    archivePrefix = "arXiv",
    primaryClass = "hep-ph",
    doi = "10.1016/j.physletb.2009.07.009",
    journal = "Phys. Lett. B",
    volume = "679",
    pages = "51--55",
    year = "2009"
}

@article{Aharony:1999ti,
    author = "Aharony, Ofer and Gubser, Steven S. and Maldacena, Juan Martin and Ooguri, Hirosi and Oz, Yaron",
    title = "{Large N field theories, string theory and gravity}",
    eprint = "hep-th/9905111",
    archivePrefix = "arXiv",
    reportNumber = "CERN-TH-99-122, HUTP-99-A027, LBNL-43113, RU-99-18, UCB-PTH-99-16, LBL-43113",
    doi = "10.1016/S0370-1573(99)00083-6",
    journal = "Phys. Rept.",
    volume = "323",
    pages = "183--386",
    year = "2000"
}

@article{Karapetyan:2018oye,
    author = "Karapetyan, G.",
    title = "{Configurational entropy and $\rho$ and $\phi$ mesons production in QCD}",
    eprint = "1802.09105",
    archivePrefix = "arXiv",
    primaryClass = "nucl-th",
    doi = "10.1016/j.physletb.2018.03.086",
    journal = "Phys. Lett. B",
    volume = "781",
    pages = "201",
    year = "2018"
}

@article{MartinContreras:2023eft,
    author = "Martin Contreras, Miguel Angel and Vega, Alfredo and Diles, Saulo",
    title = "{Isospectrality and configurational entropy as testing tools for bottom-up AdS/QCD}",
    eprint = "2308.16007",
    archivePrefix = "arXiv",
    primaryClass = "hep-ph",
    doi = "10.1016/j.physletb.2024.138723",
    journal = "Phys. Lett. B",
    volume = "854",
    pages = "138723",
    year = "2024"
}

@article{Bernardini:2016qit,
    author = "Bernardini, Alex E. and Braga, Nelson R. F. and da Rocha, Roldao",
    title = "{Configurational entropy of glueball states}",
    eprint = "1609.01258",
    archivePrefix = "arXiv",
    primaryClass = "hep-th",
    doi = "10.1016/j.physletb.2016.12.007",
    journal = "Phys. Lett. B",
    volume = "765",
    pages = "81--85",
    year = "2017"
}

@article{Karapetyan:2021ufz,
    author = "Karapetyan, G. and da Rocha, R.",
    title = "{Configurational entropy of heavy-quark QCD exotica}",
    eprint = "2103.10863",
    archivePrefix = "arXiv",
    primaryClass = "hep-ph",
    doi = "10.1140/epjp/s13360-021-01942-7",
    journal = "Eur. Phys. J. Plus",
    volume = "136",
    number = "10",
    pages = "993",
    year = "2021"
}

@article{Colangelo:2008us,
    author = "Colangelo, P. and De Fazio, F. and Giannuzzi, Floriana and Jugeau, F. and Nicotri, S.",
    title = "{Light scalar mesons in the soft-wall model of AdS/QCD}",
    eprint = "0807.1054",
    archivePrefix = "arXiv",
    primaryClass = "hep-ph",
    reportNumber = "BARI-TH-08-593",
    doi = "10.1103/PhysRevD.78.055009",
    journal = "Phys. Rev. D",
    volume = "78",
    pages = "055009",
    year = "2008"
}

@article{Rinaldi:2024fgx,
    author = "Rinaldi, Matteo and Vento, Vicente",
    title = "{Hybrid spectroscopy within the graviton soft-wall model}",
    eprint = "2402.11959",
    archivePrefix = "arXiv",
    primaryClass = "hep-ph",
    doi = "10.1103/PhysRevD.109.114030",
    journal = "Phys. Rev. D",
    volume = "109",
    number = "11",
    pages = "114030",
    year = "2024"
}

@article{Colangelo:2018mrt,
    author = "Colangelo, P. and Loparco, F.",
    title = "{Configurational Entropy can disentangle conventional hadrons from exotica}",
    eprint = "1811.05272",
    archivePrefix = "arXiv",
    primaryClass = "hep-ph",
    reportNumber = "BARI-TH/18-717",
    doi = "10.1016/j.physletb.2018.11.053",
    journal = "Phys. Lett. B",
    volume = "788",
    pages = "500--504",
    year = "2019"
}

@article{Ferreira:2020iry,
    author = "Ferreira, Luiz F. and da Rocha, Roldao",
    title = "{Nucleons and higher spin baryon resonances: An AdS/QCD configurational entropic incursion}",
    eprint = "2004.04551",
    archivePrefix = "arXiv",
    primaryClass = "hep-th",
    doi = "10.1103/PhysRevD.101.106002",
    journal = "Phys. Rev. D",
    volume = "101",
    number = "10",
    pages = "106002",
    year = "2020"
}

@article{Braga:2020opg,
    author = "Braga, Nelson R. F. and Junqueira, Octavio C.",
    title = "{Configuration entropy and confinement/deconfinement transiton in holographic QCD}",
    eprint = "2010.00714",
    archivePrefix = "arXiv",
    primaryClass = "hep-th",
    doi = "10.1016/j.physletb.2021.136082",
    journal = "Phys. Lett. B",
    volume = "814",
    pages = "136082",
    year = "2021"
}

@article{DaRold:2005mxj,
    author = "Da Rold, Leandro and Pomarol, Alex",
    title = "{Chiral symmetry breaking from five dimensional spaces}",
    eprint = "hep-ph/0501218",
    archivePrefix = "arXiv",
    reportNumber = "UAB-FT-578",
    doi = "10.1016/j.nuclphysb.2005.05.009",
    journal = "Nucl. Phys. B",
    volume = "721",
    pages = "79--97",
    year = "2005"
}

@article{Pomarol:2008aa,
    author = "Pomarol, Alex and Wulzer, Andrea",
    title = "{Baryon Physics in Holographic QCD}",
    eprint = "0807.0316",
    archivePrefix = "arXiv",
    primaryClass = "hep-ph",
    reportNumber = "UAB-FT-650",
    doi = "10.1016/j.nuclphysb.2008.10.004",
    journal = "Nucl. Phys. B",
    volume = "809",
    pages = "347--361",
    year = "2009"
}

@article{Csaki,
    author = "Csaki, Csaba and Reece, Matthew",
    title = "{Toward a systematic holographic QCD: A Braneless approach}",
    eprint = "hep-ph/0608266",
    archivePrefix = "arXiv",
    doi = "10.1088/1126-6708/2007/05/062",
    journal = "JHEP",
    volume = "05",
    pages = "062",
    year = "2007"
}

@article{FolcoCapossoli:2019imm,
    author = "Folco Capossoli, Eduardo and Mart\'\i{}n Contreras, Miguel Angel and Li, Danning and Vega, Alfredo and Boschi-Filho, Henrique",
    title = "{Hadronic spectra from deformed AdS backgrounds}",
    eprint = "1903.06269",
    archivePrefix = "arXiv",
    primaryClass = "hep-ph",
    doi = "10.1088/1674-1137/44/6/064104",
    journal = "Chin. Phys. C",
    volume = "44",
    number = "6",
    pages = "064104",
    year = "2020"
}

@article{MartinContreras:2020cyg,
    author = "Martin Contreras, Miguel Angel and Vega, Alfredo",
    title = "{Nonlinear Regge trajectories with AdS/QCD}",
    eprint = "2004.10286",
    archivePrefix = "arXiv",
    primaryClass = "hep-ph",
    doi = "10.1103/PhysRevD.102.046007",
    journal = "Phys. Rev. D",
    volume = "102",
    number = "4",
    pages = "046007",
    year = "2020"
}

@article{MartinContreras:2022lxl,
    author = "Martin Contreras, Miguel Angel and Vega, Alfredo and Diles, Saulo",
    title = "{A holographic bottom-up description of light nuclide spectroscopy and stability}",
    eprint = "2206.01834",
    archivePrefix = "arXiv",
    primaryClass = "hep-ph",
    doi = "10.1016/j.physletb.2022.137551",
    journal = "Phys. Lett. B",
    volume = "835",
    pages = "137551",
    year = "2022"
}

@article{Witten:1998qj,
    author = "Witten, Edward",
    title = "{Anti-de Sitter space and holography}",
    eprint = "hep-th/9802150",
    archivePrefix = "arXiv",
    reportNumber = "IASSNS-HEP-98-15",
    doi = "10.4310/ATMP.1998.v2.n2.a2",
    journal = "Adv. Theor. Math. Phys.",
    volume = "2",
    pages = "253--291",
    year = "1998"
}

@article{daRocha:2021xwq,
    author = "da Rocha, Roldao",
    title = "{Holographic entanglement entropy, deformed black branes, and deconfinement in AdS/QCD}",
    eprint = "2111.01244",
    archivePrefix = "arXiv",
    primaryClass = "hep-th",
    doi = "10.1103/PhysRevD.105.026014",
    journal = "Phys. Rev. D",
    volume = "105",
    number = "2",
    pages = "026014",
    year = "2022"
}

@article{Gherghetta:2009ac,
    author = "Gherghetta, Tony and Kapusta, Joseph I. and Kelley, Thomas M.",
    title = "{Chiral symmetry breaking in the soft-wall AdS/QCD model}",
    eprint = "0902.1998",
    archivePrefix = "arXiv",
    primaryClass = "hep-ph",
    doi = "10.1103/PhysRevD.79.076003",
    journal = "Phys. Rev. D",
    volume = "79",
    pages = "076003",
    year = "2009"
}

@article{Toniato:2025gts,
    author = "Toniato, Bruno and Dudal, David and Mahapatra, Subhash and da Rocha, Roldao and Jena, Siddhi Swarupa",
    title = "{Holographic QCD model for heavy and exotic mesons at finite density: A self-consistent dynamical approach}",
    eprint = "2502.12694",
    archivePrefix = "arXiv",
    primaryClass = "hep-th",
    doi = "10.1103/pn9x-ycnh",
    journal = "Phys. Rev. D",
    volume = "111",
    number = "12",
    pages = "126021",
    year = "2025"
}

@article{Dudal:2018ztm,
    author = "Dudal, David and Mahapatra, Subhash",
    title = "{Interplay between the holographic QCD phase diagram and entanglement entropy}",
    eprint = "1805.02938",
    archivePrefix = "arXiv",
    primaryClass = "hep-th",
    doi = "10.1007/JHEP07(2018)120",
    journal = "JHEP",
    volume = "07",
    pages = "120",
    year = "2018"
}

@article{Gutsche:2019pls,
    author = "Gutsche, Thomas and Lyubovitskij, Valery E. and Schmidt, Ivan and Trifonov, Andrey Yu",
    title = "{Baryons in a soft-wall AdS-Schwarzschild approach at low temperature}",
    eprint = "1905.02577",
    archivePrefix = "arXiv",
    primaryClass = "hep-ph",
    doi = "10.1103/PhysRevD.99.114023",
    journal = "Phys. Rev. D",
    volume = "99",
    number = "11",
    pages = "114023",
    year = "2019"
}

@article{Benakli:2023aes,
    author = "Benakli, Karim and Daniel, Cassiano A. and Ke, Wenqi",
    title = "{Spin-3/2 and spin-2 charged massive states in a constant electromagnetic background}",
    eprint = "2302.06630",
    archivePrefix = "arXiv",
    primaryClass = "hep-th",
    doi = "10.1007/JHEP03(2023)212",
    journal = "JHEP",
    volume = "03",
    pages = "212",
    year = "2023"
}

@article{Guo:2024nrf,
    author = "Guo, Xi and Martin Contreras, Miguel Angel and Chen, Xun and Xiang, Dong",
    title = "{Holographic bottom-up approach to {\ensuremath{\Sigma}} baryons*}",
    eprint = "2404.16608",
    archivePrefix = "arXiv",
    primaryClass = "hep-ph",
    doi = "10.1088/1674-1137/ad7d75",
    journal = "Chin. Phys. C",
    volume = "49",
    number = "1",
    pages = "013104",
    year = "2025"
}

@article{Diles:2025xot,
    author = "Diles, Saulo and Martin Contreras, Miguel Angel and Vega, Alfredo",
    title = "{Holographic quark masses and radiative decays of heavy vector mesons}",
    eprint = "2505.14324",
    archivePrefix = "arXiv",
    primaryClass = "hep-ph",
    doi = "10.1103/7dc8-kw2j",
    journal = "Phys. Rev. D",
    volume = "112",
    number = "5",
    pages = "056010",
    year = "2025"
}

@article{Maldacena:1997re,
    author = "Maldacena, Juan Martin",
    title = "{The Large N limit of superconformal field theories and supergravity}",
    eprint = "hep-th/9711200",
    archivePrefix = "arXiv",
    reportNumber = "HUTP-97-A097, HUTP-98-A097",
    doi = "10.4310/ATMP.1998.v2.n2.a1",
    journal = "Adv. Theor. Math. Phys.",
    volume = "2",
    pages = "231--252",
    year = "1998"
}

@article{Gubser:1998bc,
    author = "Gubser, S. S. and Klebanov, Igor R. and Polyakov, Alexander M.",
    title = "{Gauge theory correlators from noncritical string theory}",
    eprint = "hep-th/9802109",
    archivePrefix = "arXiv",
    reportNumber = "PUPT-1767",
    doi = "10.1016/S0370-2693(98)00377-3",
    journal = "Phys. Lett. B",
    volume = "428",
    pages = "105--114",
    year = "1998"
}

@article{Ballon-Bayona:2017sxa,
    author = "Ballon-Bayona, Alfonso and Boschi-Filho, Henrique and Mamani, Luis A. H. and Miranda, Alex S. and Zanchin, Vilson T.",
    title = "{Effective holographic models for QCD: glueball spectrum and trace anomaly}",
    eprint = "1708.08968",
    archivePrefix = "arXiv",
    primaryClass = "hep-th",
    doi = "10.1103/PhysRevD.97.046001",
    journal = "Phys. Rev. D",
    volume = "97",
    number = "4",
    pages = "046001",
    year = "2018"
}

@article{Hong:2006ta,
    author = "Hong, Deog Ki and Inami, Takeo and Yee, Ho-Ung",
    title = "{Baryons in AdS/QCD}",
    eprint = "hep-ph/0609270",
    archivePrefix = "arXiv",
    reportNumber = "PNUTP-06-A20, KIAS-P06040",
    doi = "10.1016/j.physletb.2007.01.030",
    journal = "Phys. Lett. B",
    volume = "646",
    pages = "165--171",
    year = "2007"
}

@article{Ahn:2009px,
    author = "Ahn, Hyo Chul and Hong, Deog Ki and Park, Cheonsoo and Siwach, Sanjay",
    title = "{Spin 3/2 Baryons and Form Factors in AdS/QCD}",
    eprint = "0904.3731",
    archivePrefix = "arXiv",
    primaryClass = "hep-ph",
    reportNumber = "PNUTP-09-A02",
    doi = "10.1103/PhysRevD.80.054001",
    journal = "Phys. Rev. D",
    volume = "80",
    pages = "054001",
    year = "2009"
}

@article{Fang:2016uer,
    author = "Fang, Zhen and Li, Danning and Wu, Yue-Liang",
    title = "{IR-improved Soft-wall AdS/QCD Model for Baryons}",
    eprint = "1602.00379",
    archivePrefix = "arXiv",
    primaryClass = "hep-ph",
    doi = "10.1016/j.physletb.2016.01.045",
    journal = "Phys. Lett. B",
    volume = "754",
    pages = "343--348",
    year = "2016"
}

@article{daRocha:2025gcz,
    author = "da Rocha, R. and Silva, P. H. O.",
    title = "{Higher-spin light-flavor baryonic spectroscopy in AdS/QCD at finite temperature}",
    eprint = "2510.05369",
    archivePrefix = "arXiv",
    primaryClass = "hep-ph",
    doi = "10.1016/j.physletb.2025.139917",
    journal = "Phys. Lett. B",
    volume = "870",
    pages = "139917",
    year = "2025"
}

@article{Wang:2015osq,
    author = "Wang, Zhiyuan and Ma, Bo-Qiang",
    title = "{A unified approach to hadron phenomenology at zero and finite temperatures in a hard-wall AdS/QCD model}",
    eprint = "1512.01957",
    archivePrefix = "arXiv",
    primaryClass = "hep-ph",
    doi = "10.1140/epja/i2016-16122-2",
    journal = "Eur. Phys. J. A",
    volume = "52",
    number = "5",
    pages = "122",
    year = "2016"
}

@article{Erlich:2005qh,
    author = "Erlich, Joshua and Katz, Emanuel and Son, Dam T. and Stephanov, Mikhail A.",
    title = "{QCD and a holographic model of hadrons}",
    eprint = "hep-ph/0501128",
    archivePrefix = "arXiv",
    reportNumber = "SLAC-PUB-10965, WM-05-101, INT-PUB-05-02",
    doi = "10.1103/PhysRevLett.95.261602",
    journal = "Phys. Rev. Lett.",
    volume = "95",
    pages = "261602",
    year = "2005"
}
\end{document}